\documentclass[10pt,aps,prd,twocolumn,showpacs,superscriptaddress,nofootinbib,nobibnotes,longbibliography,floatfix]{revtex4-1}

\usepackage{amsmath}
\usepackage{graphicx} 
\usepackage[utf8]{inputenc}
\usepackage[T1]{fontenc}
\usepackage{hhline}
\usepackage{enumitem}
\usepackage{bm}
\usepackage{mathtools,amsmath,amssymb,amsfonts,graphicx,tensor,csquotes,accents, commath,chngcntr,siunitx}
\usepackage[dvipsnames]{xcolor}
\usepackage{hyperref}
\hypersetup{colorlinks=true, citecolor=MidnightBlue,
            linkcolor=MidnightBlue, urlcolor=MidnightBlue, linktocpage=true}
\usepackage[normalem]{ulem}
\usepackage{tensor}
\usepackage{leftindex}
\usepackage{caption}
\usepackage{subcaption}
\usepackage{amssymb} 
\usepackage{orcidlink}
\usepackage{pifont}
\usepackage{lmodern}
\usepackage{cancel}

\usepackage{booktabs}
\usepackage{float}

\usepackage{array}

\usepackage{xcolor}

\usepackage{tabularray}

\DeclareMathAlphabet{\mathpzc}{OT1}{pzc}{m}{it}
\definecolor{darkgreen}{rgb}{0.0, 0.6, 0.0}

\begin{document}

\title{Dynamical hair growth in black hole binaries in Einstein-scalar-Gauss-Bonnet gravity}

\author{Lodovico Capuano \orcidlink{0009-0001-0369-6635}}
\email{lcapuano@sissa.it}
\email{lodovico.capuano@uniroma1.it}
\affiliation{SISSA, Via Bonomea 265, 34136 Trieste, Italy \& INFN Sezione di Trieste}
\affiliation{IFPU - Institute for Fundamental Physics of the Universe, Via Beirut 2, 34014 Trieste, Italy}
\affiliation{Dipartimento di Fisica, Sapienza Università di Roma, Piazzale Aldo Moro 5, 00185, Roma, Italy}

    \author{Llibert Arest\'e Sal\'o \orcidlink{0000-0002-3812-8523}}
	\email{llibert.arestesalo@kuleuven.be}
	\affiliation{Instituut voor Theoretische Fysica, KU Leuven. Celestijnenlaan 200D, B-3001 Leuven, Belgium}
        \affiliation{Leuven Gravity Institute, KU Leuven. Celestijnenlaan 200D, B-3001 Leuven, Belgium}
        
        \author{Daniela D. Doneva \orcidlink{0000-0001-6519-000X}}
	\email{daniela.doneva@uni-tuebingen.de}
        \affiliation{Departament d'Astronomia i Astrof\'isica, Universitat de Val\`encia, Dr. Moliner 50, 46100, Burjassot (Val\`encia), Spain}
	\affiliation{Theoretical Astrophysics, Eberhard Karls University of T\"ubingen, T\"ubingen 72076, Germany}

	\author{Stoytcho S. Yazadjiev \orcidlink{0000-0002-1280-9013}}
	\email{yazad@phys.uni-sofia.bg}
	\affiliation{Department of Theoretical Physics, Faculty of Physics, Sofia University, Sofia 1164, Bulgaria}
	\affiliation{Institute of Mathematics and Informatics, Bulgarian Academy of Sciences, Acad. G. Bonchev St. 8, Sofia 1113, Bulgaria}
    \author{Enrico Barausse \orcidlink{0000-0001-6499-6263}}
\email{barausse@sissa.it}
\affiliation{SISSA, Via Bonomea 265, 34136 Trieste, Italy \& INFN Sezione di Trieste}
\affiliation{IFPU - Institute for Fundamental Physics of the Universe, Via Beirut 2, 34014 Trieste, Italy}

\begin{abstract}
Within the framework of scalar–tensor theories of gravity, certain models can evade classical black hole no-hair theorems. A well-known example is Einstein–scalar–Gauss–Bonnet gravity, where black holes carrying a scalar charge can exist. We find 
that, within this theory, binary black holes   initially described by General Relativity
can acquire scalar charges once they reach a critical orbital separation (``dynamical scalarization'').
We develop a simple semi-analytic model, based on the adiabatic conservation of the total Wald entropy, to estimate the scalar charge evolution during the binary inspiral. We also run fully nonlinear numerical-relativity simulations for different configurations, finding consistent results.
The gravitational-wave phase difference between Einstein-scalar-Gauss-Bonnet and General Relativity waveforms, which we use to assess detectability, is also computed. We find that dynamical scalarization
might  be observable in nearly equal-mass binary black hole mergers  with third-generation ground-based gravitational-wave detectors, in a narrow range of the dimensional coupling of the theory.
\end{abstract}

\maketitle

\section{Introduction}
General Relativity (GR) represents the standard framework for describing gravitational phenomena, and its predictions have been extensively tested in a wide range of experiments and observations~\cite{Will_2014,Shapiro:1964uw,Einstein:1915bz,Einstein:1911vc,Dyson:1920cwa}. In particular, gravitational waves (GWs) emitted during the inspiral and merger of compact binary systems have opened an unprecedented observational window on the strong-field regime of gravity~\cite{LIGOScientific:2016aoc,LIGOScientific:2016lio,LIGOScientific:2020ibl,LIGOScientific:2021sio,KAGRA:2021duu,KAGRA:2021vkt,LIGOScientific:2021usb,LIGOScientific:2021qlt,LIGOScientific:2025slb,Akyuz:2025seg}. So far, all detections are consistent with the predictions of GR, with no statistically significant evidence for deviations~\cite{LIGOScientific:2021sio,Wang:2024erh,KAGRA:2025oiz,LIGOScientific:2025obp}.

However, GR presents issues on both the theoretical and phenomenological side. In fact, despite being a good effective field theory (EFT), it still lacks a coherent ultraviolet (UV) completion. The incompleteness of GR can be seen, for instance, in the presence of singular solutions, such as black holes (BHs)~\cite{Penrose:1964wq,Hawking:1973uf,Landsman:2022hrn}. On the phenomenological side, the dynamics of the universe on cosmological scales appears to require either a dark-energy/cosmological-constant component, or an infrared (IR) extension of GR~\cite{Planck:2018vyg,SupernovaCosmologyProject:1998vns,Riess1998}. It is therefore crucial to extend current tests of gravity to stronger and more dynamical regimes.

Several classical extensions of GR have been proposed in the last decades.
Among these, one of the oldest and simplest is the class of scalar-tensor (ST) theories,
where a scalar degree of freedom is present alongside the tensor graviton. The simplest theory in this class is
 Fierz-Jordan-Brans-Dicke theory~\cite{Fierz:1939ix,Jordan:1959eg,Brans:1961sx,Dicke:1961gz}, in which the scalar field is conformally coupled (linearly) to matter, but no higher-derivative interactions or couplings with higher curvature invariants are present. This kind of model has also been extended to non-linear conformal couplings~\cite{Damour:1992we,Damour:1993hw}. 
More general ST theories, including all possible interaction terms that lead to second-order equations of motion, are known as Horndeski theories and were developed later~\cite{Horndeski:1974wa,Gleyzes:2014dya,Langlois:2017mdk}.
The motivation for this kind of classical modifications of gravity is mostly related to the IR regime. In fact, it has been shown that some specific models of ST gravity are able to provide self-accelerating solutions 
that may provide a dark-energy-like cosmological phenomenology~\cite{Chiba:1999ka,Arkani-Hamed:2003pdi,Nicolis:2008in}. However, although ST gravity does not directly address the UV problems of GR, the presence of at least one scalar (or pseudo-scalar) degree of freedom at low energies is generally expected in UV-complete frameworks like string theory~\cite{Becker:2006dvp}. 

No-hair theorems imply that most  interaction terms in ST gravity do not produce deviations from the stationary BH solutions of GR~\cite{Bekenstein:1971hc, Bekenstein:1972ky, Hui:2012qt, Graham:2014mda, Herdeiro:2015waa, Capuano:2023yyh, Yazadjiev:2025ezx}. In other words, the scalar field settles into a trivial configuration and does not provide an additional charge to the BH. A possible exception~\cite{Sotiriou:2013qea,Sotiriou:2014pfa} arises if the scalar field is coupled to the  Gauss-Bonnet (GB) invariant,\footnote{Another possibility is to partially drop the assumption of stationarity and allow the scalar field to have a linear dependence on time. This works in shift-symmetric theories, where the field equations only depend on derivatives of the scalar, making the linear time dependence disappear~\cite{Capuano:2023yyh}.} defined as
\begin{equation}
   \mathcal{G}=R^{\mu\nu\rho\sigma}R_{\mu\nu\rho\sigma}-4 R^{\mu\nu}R_{\mu\nu}+R^2\,,
\end{equation}
where $R_{\mu\nu\rho\sigma}$, $R_{\mu\nu}$ and $R$ are, respectively, the Riemann tensor, the Ricci tensor, and the Ricci scalar. The subset of ST gravity featuring a standard kinetic term for the scalar and a coupling with $\mathcal{G}$ will hereafter be referred to as Einstein-scalar-GB (EsGB) gravity.

The most stringent limits on the coupling constant $\lambda$
between the scalar and the Gauss-Bonnet invariant arise from the observation of relatively light BHs or neutron stars \cite{Danchev:2021tew,Wong:2022wni,Lyu:2022gdr,Yordanov:2024lfk}. This behavior is a direct consequence of the fact that $\lambda$ carries physical dimensions: the smaller the characteristic scale of the BH, and thus the larger the curvature at the BH horizon, the tighter the bound becomes. From an EFT viewpoint, interaction terms with a dimensional coupling represent irrelevant operators, suppressed at large scales (or low energies) compared to $\lambda$. Therefore, for $\lambda$ in a suitable range, supermassive BHs are effectively described by GR in EsGB,  while stellar-origin BHs 
would show deviations from GR.\footnote{In more complicated models where a coupling to $\mathcal{G}^2$ is introduced,  one can construct
theories where supermassive BHs have scalar charges, but stellar-origin BHs do not~\cite{Eichhorn:2023iab,Smarra:2025syw}. However, it has been argued that these models are not consistent EFTs~\cite{Thaalba:2025ljh}. In this work, we do not explore these theories.}

A particularly interesting case is represented by ST theories that admit both GR  and hairy solutions, with the latter appearing only for systems with larger/smaller curvature at the BH horizon. In this case, compact objects can experience phase transitions from one branch of solutions to the other, undergoing scalarization/descalarization~\cite{Doneva:2017bvd,Silva:2017uqg,Doneva2022BeyondTS,Herdeiro:2020wei,Dima:2020yac}. This process can be triggered in several possible ways, both in isolation and in binary systems (see~\cite{doneva2022scalarization} for a review). The latter case is the dynamical (de)scalarization (DS) mechanism, which occurs when the effective compactness of the binary system exceeds a certain threshold, while the individual objects cannot carry a scalar charge when isolated. In what follows, the critical (de)scalarization separation will be referred to as the (de)scalarization radius. This process could  impact the inspiral dynamics, possibly leading to an observable signature in the GW signal~\cite{Sampson_2014}. DS was first discovered in the context of binary neutron stars~\cite{Barausse_2013, Shibata:2013pra,Palenzuela_2014,Taniguchi:2014fqa,Sampson_2014,Sennett:2016rwa,Sennett:2017lcx,Khalil:2019wyy,Kuan:2023trn}, but  it has recently been shown to also affect BH binaries in EsGB gravity~\cite{Julie:2023ncq}.

Until very recently, these theories had only been studied with the post-Newtonian (PN) formalism~\cite{Shiralilou:2020gah,Shiralilou:2021mfl,Julie:2019sab,Yagi:2011xp,Lyu:2022gdr} and with numerical relativity (NR) simulations in the decoupling approximation~\cite{Okounkova:2017yby,Silva:2020omi,Elley:2022ept,Doneva:2022byd,Evstafyeva:2022rve} (in which the scalar evolves nonlinearly, but does not back-react onto the metric). These limitations were overcome by the introduction of the modified harmonic gauge by Kov\'acs and Reall, which led to a well-posed formulation of the Horndeski and Lovelock families of theories in the weak coupling regime~\cite{Kovacs:2020pns,Kovacs:2020ywu}, which includes EsGB gravity in the EFT regime. This rapidly led to the first  nonlinear NR simulations in EsGB gravity, both in harmonic coordinates~\cite{East:2020hgw,East:2021bqk,East:2022rqi} and in the moving-punctures approach~\cite{AresteSalo:2022hua,AresteSalo:2023mmd,AresteSalo:2023hcp}. Although these studies are still in an early stage, recent results have shown consistency with PN calculations~\cite{Corman:2022xqg,Corman:2024vlk,AresteSalo:2025sxc} and some novel signatures have already been found~\cite{Lara:2025kzj,Corman:2025wun}.

In this work, we study the DS of BH binaries in EsGB theory. We first find evidence of this process with a semi-analytic approach that allows for tracking the scalar field at the BH event horizon as a function of separation, assuming that the inspiral is  adiabatic. Moreover, we perform the first fully nonlinear NR simulations in EsGB gravity that show DS in binary BH (BBH) systems. We find these numerical results to be consistent with our semi-analytic approach in its regime of validity. We finally comment on the detectability of DS with third generation ground-based GW detectors, such as ET.

In Sec.~\ref{sec2} we describe the EsGB model that we consider,
 and we introduce our semi-analytic approach  to computing the scalar field at the BH event horizon as a function of the binary separation. In Sec.~\ref{sec3} we describe the details of the NR simulations and discuss their results, also in comparison with the semi-analytic model. In Sec.~\ref{sec4}, we compute semi-analytically the dephasing between GWs produced in GR and in EsGB. Finally, in Sec.~\ref{sec5}, we address the detectability of BH DS with next-generation GW detectors.
Throughout this paper, we  work in geometric units $c = G = 1$, and in the  mostly plus signature of the metric $(-,+,+,+)$.

\section{The model}
\label{sec2}
Before describing the details of EsGB theory, let us introduce a semi-analytic description of a binary system of compact objects carrying scalar charges. In full generality, one can define the scalar charge $Q$, in analogy with the electrostatic charge, as the coefficient of the leading term (besides a possible constant background) in a large-distance expansion of the scalar field sourced by the object itself. In other words, the scalar field at a (large) distance $d$ from the objects reads
\begin{equation}
    \varphi(d)=\varphi_\infty+\frac{Q}{r}+\mathcal{O}\left(\frac{1}{r}\right)^{2}\,.
    \label{eq:large_d_exp}
\end{equation}
For nonrotating compact objects, the scalar charge is typically a function of two variables, i.e.  the background value of the scalar field $\varphi_\infty$ and another parameter related to the characteristic mass/energy scale of the object. In the case of BHs, this latter parameter can be chosen to be the entropy, that we will indicate with $\mathcal{S}$. Thus, in general, the scalar charge will be described by the function $Q(\mathcal{S},\varphi_{\infty})$. The background scalar field $\varphi_\infty$ can be sourced by cosmological dynamics or by a far-away compact companion. In the following, we will assume no contribution from cosmological scalar fields, and focus on the case in which the background scalar field $\varphi_\infty$ is sourced by a second object. 
\begin{figure}[t]
          \centering
\includegraphics[width =0.495\textwidth]{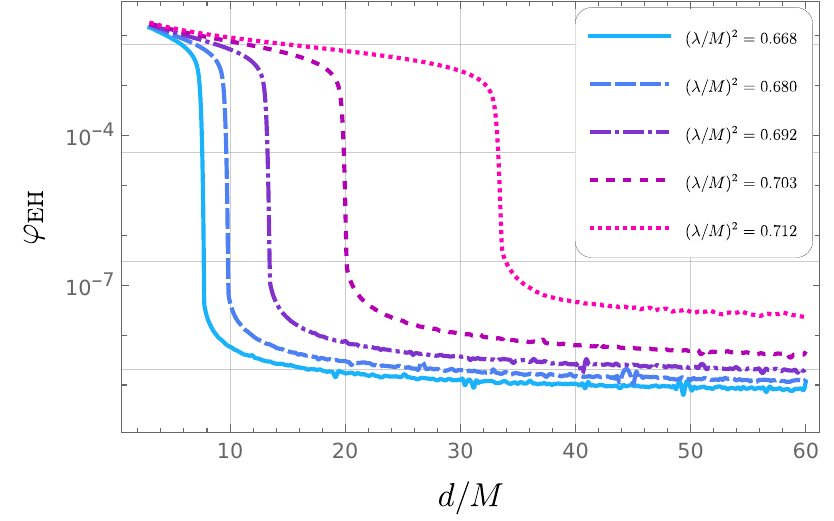}
 \caption{Scalar field at the event horizon $\varphi_{\rm EH}$ as a function of separation for $\beta = 800$ and different values of $\lambda$, computed with the semi-analytic approach described in the main text. The scalarization radius increases for larger values of the coupling. 
}
\label{fig:DS_different_lambda2}
\end{figure}
Let us now go a step further and consider two compact objects $A$ and $B$ in a binary system. The presence of the companion $B$ affects the scalar charge of the compact object $A$ and vice-versa. This mutual interaction between the two scalar hairs is in principle a highly nonperturbative phenomenon. However, it can be efficiently captured, at leading order, by a simple model, which was introduced in~\cite{Palenzuela_2014} for binary neutron star systems (see also Ref.~\cite{Sennett:2016rwa}). A similar procedure is carried out in~\cite{Julie:2023ncq} in the context of BBHs. We provide a comparison between the two approaches in Appendix~\ref{Comparison_Juliè}.

The main idea of our method is to generalize Eq.~\eqref{eq:large_d_exp} to a two-body configuration, i.e. to write down the following system of coupled algebraic equations
\begin{equation}
    \begin{split}
        &\varphi_A=\frac{Q(\mathcal{S},\varphi_B)}{d}\,,\\
        &\varphi_B=\frac{Q(\mathcal{S},\varphi_A)}{d}\,,
    \end{split}
    \label{algebraic_system}
\end{equation}
where the $r$ of Eq.~\eqref{eq:large_d_exp} is now replaced by the separation between $A$ and $B$, indicated by $d$.  Note that the expansion in $1/d$ can be truncated at leading order for our purposes. The background scalar field felt by A, which we previously denoted by $\varphi_\infty$, is now given by the field produced by the companion object B, and vice-versa. One can search for a solution of the system of Eq.~\eqref{algebraic_system} iteratively. We are interested in the case in which two BHs without scalar charge in isolation acquire it in a binary system, which is the definition of DS~\cite{Palenzuela_2014}. The existence of a solution to Eq.~\eqref{algebraic_system} with non-vanishing scalar charges, at fixed separation $d$, would signal DS. To practically carry out this procedure, one  has to first obtain a function $Q(\mathcal{S},\varphi_\infty)$, which depends on the specific theory of gravity. The total entropy of a BBH system, in principle should change as $d$ evolves. However, the inspiral can be modelled, as a first approximation, as an adiabatic process. For this reason, we will assume that the total entropy is fixed at different separations.

Let us now consider the EsGB action 
\begin{equation}
    S=\frac{1}{16\pi }\int {\rm d}^4x\,\sqrt{-g}\left[R-2\partial_\mu \varphi\partial^\mu\varphi +\lambda^2 f(\varphi)\mathcal{G} \right]\,,
\label{eq:action}
\end{equation}
where $\varphi$ is the scalar degree of freedom, $f(\varphi)$ the coupling function, and $\lambda$ the coupling constant, which is the characteristic length scale of the theory. In this work we will consider $\lambda$ to be of the order $\sim \mathcal{O}(10-100)\,{\rm km}$, which is the scale for which the inspiral dynamics of compact objects can be affected by the scalar-GB coupling. On the other hand, this is consistent with the choice of neglecting any non-trivial cosmological background for the scalar field at larger scales, i.e. $\dot\varphi\to 0$ asymptotically. This also ensures that the model we are considering is consistent with observational bounds from the GW propagation speed~\cite{Creminelli:2017sry,LIGOScientific:2017vwq,Gong:2017kim}. In this work, we will consider the coupling function
\begin{equation}
    f(\varphi)=\frac{1}{2\beta}\left(1-e^{-\beta \varphi^2}\right)\,,
\label{eq:coupling}
\end{equation}
where $\beta$ is an arbitrary dimensionless constant.We note that DS with this coupling function was also briefly studied in~\cite{Julie:2023ncq}. This choice is motivated by the fact that it allows for spontaneous scalarization of BHs, and the resulting hairy solutions can be linearly stable~\cite{Doneva2022BeyondTS}. That does not happen for simpler couplings, e.g. quadratic in $\varphi$~\cite{Blazquez-Salcedo:2018jnn,Silva:2020omi}.
\begin{figure}[t]
          \centering
\includegraphics[width =0.48\textwidth]{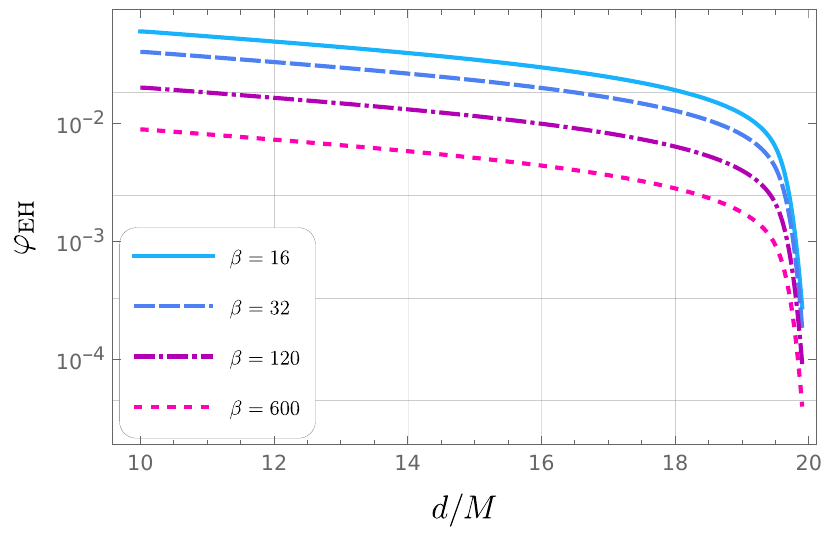}
 \caption{Scalar field at the event horizon (EH) as a function of separation for $(\lambda/M)^2 = 0.703$ and different values of $\beta$, computed with the semi-analytic approach described in the main text. The scalarization radius is constant, while the scalar charge increases for decreasing $\beta$.
}
\label{fig:DS_different_beta}
\end{figure}
Moreover, the fact that this coupling function is not shift-symmetric ensures that different $\varphi_\infty$ represent solutions that are physically distinct. This would not be the case e.g. for the choice $f(\varphi)\sim \varphi$. We also stress the fact that the coupling of Eq.~\eqref{eq:coupling} is not motivated as the low-energy limit of UV complete quantum gravity theories~\cite{Becker:2006dvp}, but is mostly justified by phenomenology at astrophysical scales.

We obtain spherically symmetric BH solutions by solving the equation of motion (see Appendix~\ref{Appendix:EOM}) with a shooting method, fixing the value of a scalar field at infinity $\varphi_{\infty}$, in a manner similar to~\cite{Doneva:2017bvd}.

As anticipated, we model the inspiral as a quasistatic process, and we thus need sequences of solutions at constant entropy, for different values of $\varphi_\infty$, as done in ~\cite{Julie:2023ncq} (see \cite{Julie:2017rpw,Cardenas:2017chu,Julie:2019sab,Julie:2022huo} for further details on the fixing of the entropy). In EsGB gravity, the Wald entropy of a static BH is given by~\cite{Wald:1993nt,Iyer:1994ys,Julie:2019sab}
\begin{equation}
    \mathcal{S}_{\rm W}=\frac{A_{\rm EH}}{4}+4\pi\lambda^2 f(\varphi_{\rm EH})\,,
\end{equation}
where $A_{\rm EH}$ and $\varphi_{\rm EH}$ are, respectively, the area of the event horizon and the scalar field at that location.

Interpolating  sequences of solutions with constant entropy, we obtain the numerical function $Q(\mathcal{S_{\rm W}},\varphi_\infty)$, which we use to iteratively solve the algebraic system given by Eq.~\eqref{algebraic_system}. 
Notice that the scalar charge represents a physical observable of the system, from which $\varphi_\infty$ can in principle be reconstructed once the entropy/mass is known. This further clarifies the physical interpretation of the asymptotic value of the scalar field in our model.

Typically, the scalar charge around the BHs will slowly increase with the increase of $\varphi_\infty$. However, for the coupling \eqref{eq:coupling}, we find a rapid increase in the scalar field at some critical distance between the two BHs. This signals the presence of a DS effect.

\begin{figure}[t]
          \centering
\includegraphics[width =0.48\textwidth]{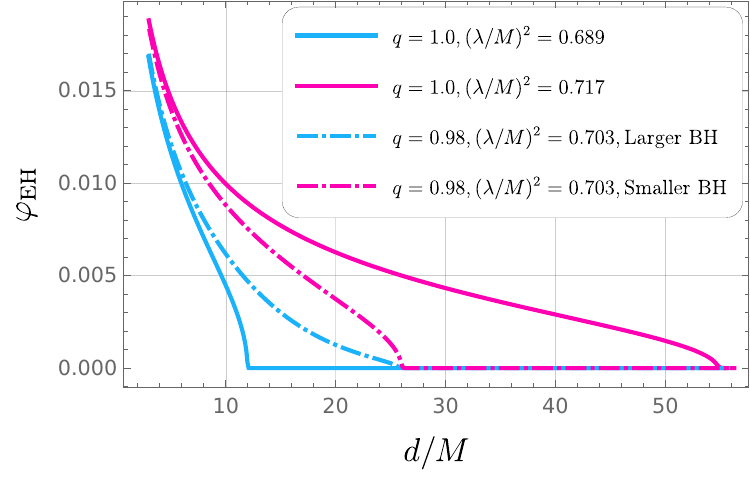}
 \caption{Effect of a slight mass asymmetry ($m_1/m_2 = 0.98$) on the scalar field evolution. Dot-dashed lines show the unequal-mass case, while solid lines correspond to equal-mass binaries with individual masses $m_1$ and $m_2$. The scalarization radius is the same for both BHs and lies between the equal-mass cases, though the scalar field growth differs at each horizon. 
}
\label{fig:q098}
\end{figure}
In  Fig.~\ref{fig:DS_different_lambda2}, we show the behavior of the scalar field at the BH event horizon, as a function of separation, normalized by the total ADM mass of the system $M$~\cite{Arnowitt:1962hi}, for  equal-mass binaries. It can clearly be observed that an increase in $\lambda/M$ leads to a larger scalarization radius $d_{\rm DS}$. The relation $d_{\rm DS}(\lambda/M)$ is highly nonlinear, and the scalarization radius diverges as we approach the limiting value $(\lambda/M)^2_{\rm crit}\equiv0.7255$. For larger values of the coupling, the scalarization radius is infinite, meaning that BHs are already scalarized in isolation. If one considers two different component masses $m_1$, $m_2$, such that $(\lambda/(2m_1))^2\,,\,(\lambda/(2m_2))^2<(\lambda/M)^2_{\rm crit}$, DS occurs at an intermediate location between the scalarization radii corresponding to mass-symmetric systems with both masses equal to $m_1$ and $m_2$, respectively. This means that, if the two BHs lie in the mass window for which DS can occur, the scalarization radius is typically maximized by equal-mass systems close to the threshold $(\lambda/M)^2_{\rm crit}$. 

For systems in which only one of the two BHs is already scalarized in isolation, \emph{induced scalarization}~\cite{Palenzuela_2014} takes place, which we do not consider in the present work. On the other hand, the parameter $\beta$ does not affect the scalarization radius, but rather the scalar charge excited in the process. This can be clearly observed in  Fig.~\ref{fig:DS_different_beta}, where we show the scalar field at the event horizon as a function of the orbital separation for $(\lambda/M)^2=0.703$ and different values of $\beta$. While DS always occurs at $d \simeq 19.75\,M$, the endpoint of the scalar field growth is larger for smaller values of $\beta$.

In  Fig.~\ref{fig:q098}, we show the effect of a mass ratio $q =m_1/m_2 \neq 1$. The dot-dashed lines repict the case of $q =0.98$, with $(\lambda/M)^2=0.703$. The magenta one represents the scalar field at the event horizon of the larger BH, while the blue one indicates the same quantity at the event horizon of the smaller BH. The continuous lines represent instead an equal mass system in which the two BH masses are equal to $m_1$ (magenta) and $m_2$ (blue). In these cases, the total mass of the system is respectively smaller and larger, implying different values of $\lambda/M$. As anticipated, one can notice that in the $q=0.98$ case, the scalarization radius lies between those corresponding to $q = 1$. Moreover, this scalarization radius is \emph{the same} for the two BHs, although the growth of the scalar field at the two horizons is different. In particular, the scalar field at the event horizon of the larger BH (blue dot-dashed line) 
approaches the scalar field at the event horizon of one component of the $q = 1$ system in which $\lambda/M$ is smaller (blue line). On the other hand, the smaller BH in the $q = 0.98$ system exhibits a scalar field profile that  gets closer to the one of the system with $q =1 $ and larger $\lambda/M$.


\section{Numerical relativity simulations}
\label{sec3}
The semi-analytic treatment described in the previous section is simple but powerful in capturing the leading-order effects related to DS. Inevitably, though, close to the merger, which is the true nonlinear part of the evolution, the semi-analytic approximation is expected to fail. Moreover, it relies on an important assumption, namely that the evolution is adiabatic, meaning that the timescale for the scalar field development is much shorter than the characteristic time of the inspiral. As we will see below, this can easily be violated for certain ranges of parameters. All this motivates the study of the problem through NR simulations of the merger, which can provide the full picture of scalar field development. This will be the focus of the current section.

\subsection{Numerical setup}
For all the simulations in this work, which feature equal-mass BBHs, we have used the \texttt{GRFolres} code~\cite{AresteSalo:2023hcp}, which is an extension of the NR code \texttt{GRChombo}~\cite{Andrade:2021rbd}. We have chosen initial data such that the masses of the individual BHs are $m_1=m_2=0.5\,M$, where $M$ is an arbitrary code unit  coinciding with the total ADM mass of the system. We use a computational domain of $L=1024\,M$ and $N=160$ grid points on the coarsest level (which we change to $N=128$ or $N=192$ when running at low resolution or high resolution, respectively), with $10$ levels of refinement with a refinement ratio of $2:1$, resulting in a finest resolution of $dx=M/80$, giving $\sim 80$ grid points across the BH horizon. We employ adaptive mesh refinement (AMR), using the tagging criterion explained in Section 4 of~\cite{Radia:2021smk}. The CCZ4 damping parameters are fixed at $\kappa_1=0.6/M$ and $\kappa_2=-0.15$, the Kreiss-Oliger numerical dissipation coefficient is set to $\sigma = 0.5$ (see~\cite{Radia:2021smk}). The modified CCZ4 parameters that we employ are $a_0=0.1$ and $b_0=0.2$ (see~\cite{AresteSalo:2022hua,AresteSalo:2023mmd} for further details). Note that these values are slightly different from the ones used in~\cite{AresteSalo:2025sxc} and in all our previous works, since we have verified that they minimize the numerical change of the mass of the system, while still allowing for stable long-term evolution. This is particularly important for our studies because DS is very sensitive to changes in the system's mass. The convergence of the code is discussed in Appendix~\ref{Convergence}.

\subsection{Initial data}

The initial data has been constructed using the \texttt{TwoPunctures} spectral solver~\cite{Ansorg:2004ds}, which provides binary puncture data of Bowen-York type and is integrated into \texttt{GRChombo}. We have computed the initial momenta of the punctures at 3PN order following~\cite{Bruegmann:2006ulg}. The list of simulations presented in this paper is given in Table \ref{tab:sim_params}.

 \begin{table}[t]
  \centering
  \setlength{\tabcolsep}{6pt}

  \caption{The simulations presented in the paper, including the coupling parameters $(\lambda/M)^2$ and $\beta$, the initial BH separation $d/M$, and the initial pulse $P_y$ (note that $P_x=0$ and $P_z=0$ for all simulations). The ID column indicates the type of scalar field initial data that was used. The column $d_{\rm DS}/M$ gives the DS radius based on semi-analytic calculations.}
\begin{tabular}{c|cccccc}
 & $(\lambda/M)^2$ & $\beta$ & $d/M$ & $P_y/M$ & ID & $d_{\rm DS}/M$ \\
\hline
(i)   & 0.688 & 48  & 11 & 0.090189 & ID1 & 11.71 \\
(ii)  & 0.688 & 48  & 13 & 0.080744 & ID1 & 11.71 \\
(iii) & 0.712 & 800 & 13 & 0.080744 & ID1 & 33.25 \\
(iv)  & 0.712 & 800 & 13 & 0.080744 & ID2 & 33.25 \\
(v)   & 0.703 & 32  & 15 & 0.080744 & ID2 & 19.75 \\
(vi)  & 0.703 & 16  & 15 & 0.080744 & ID2 & 19.75 \\
(vii) & 0 (GR) & -- & 15 & 0.080744 & ID2 & 19.75 \\
(viii) & 0.703 & 48 & 13 & 0.080744 & ID2 & 19.75 \\
\hline
\end{tabular}
\label{tab:sim_params}
\end{table}

In the case of DS with a coupling function of the form of Eq.~\eqref{eq:coupling}, GR initial data satisfy the constraint equations and represent a quasi-stationary equilibrium, because in this theory $\varphi=0$ is a solution of the field equations. When the criteria for DS are fulfilled, the GR quasi-stationary equilibrium solution is no longer stable, and thus scalar hair develops. To trigger this nonlinear scalar field development during the numerical simulations, we have to add a small seed scalar field. Since the growth of the scalar field for the simulations we have performed is typically slow compared to the timescale of the merger, we have used two types of scalar field initial data, to check the consistency of our results independently of the initial data. 

In the first case (ID1), we add a constant scalar field with an amplitude of $10^{-5}$. In the second case (ID2), we add a scalar field with a Gaussian profile around the two BHs, with a width of $1M$ and an amplitude of $10^{-2}$. We note that such scalar field initial data violates the constraints by a small amount. Ideally, one would start with self-consistent quasi-stationary initial data within EsGB with a nonzero scalar field. Unfortunately, even though there are important developments in this direction~\cite{Brady:2023dgu,Nee:2024bur}, such initial data are still  unavailable. This is why we rely on the constraint damping terms in the equations of motion to eliminate these constraint vioations during the evolution. Indeed, we have verified that on a timescale of less than $400\,M$, the constraint violations due to the initial scalar field configuration, as well as the initial gauge settling, are suppressed. This is demonstrated in Fig.~\ref{fig:ComparePhiID} (bottom panel), which will be analyzed in detail in the next subsection. Moreover, we have verified that the scalar field evolution, the eccentricity of the system, as well as the GW dephasing, are largely unaffected by the exact choice of the initial scalar field data for the theory parameters considered in the present paper. 

Given that we start from a separation between $11\,M$ and $15\,M$, the runs provide roughly 8 to 19 orbits before merger. The eccentricity is relatively low, e.g., attaining roughly a maximum of $0.006$ among all models in Table \ref{tab:sim_params}. Furthermore, we have verified that for the simulations performed in the present paper, DS does not lead to a noticeable eccentricity increase. 

\subsection{Scalar field development in merger simulations}
\begin{figure}
\centering
\includegraphics[width =0.48\textwidth]{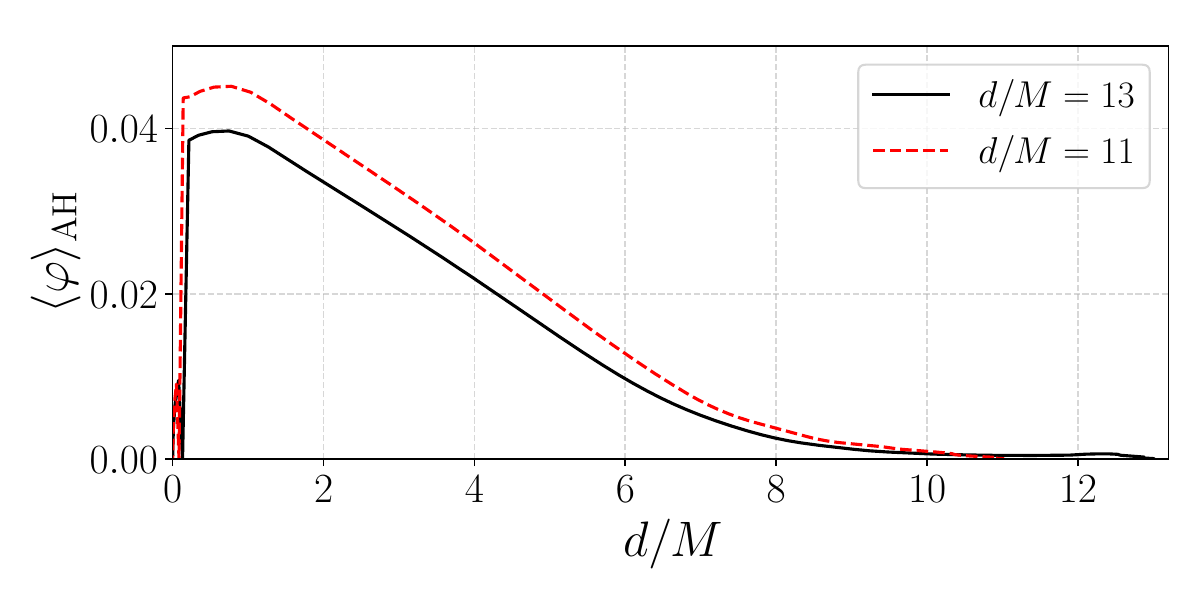}
\includegraphics[width =0.48\textwidth]{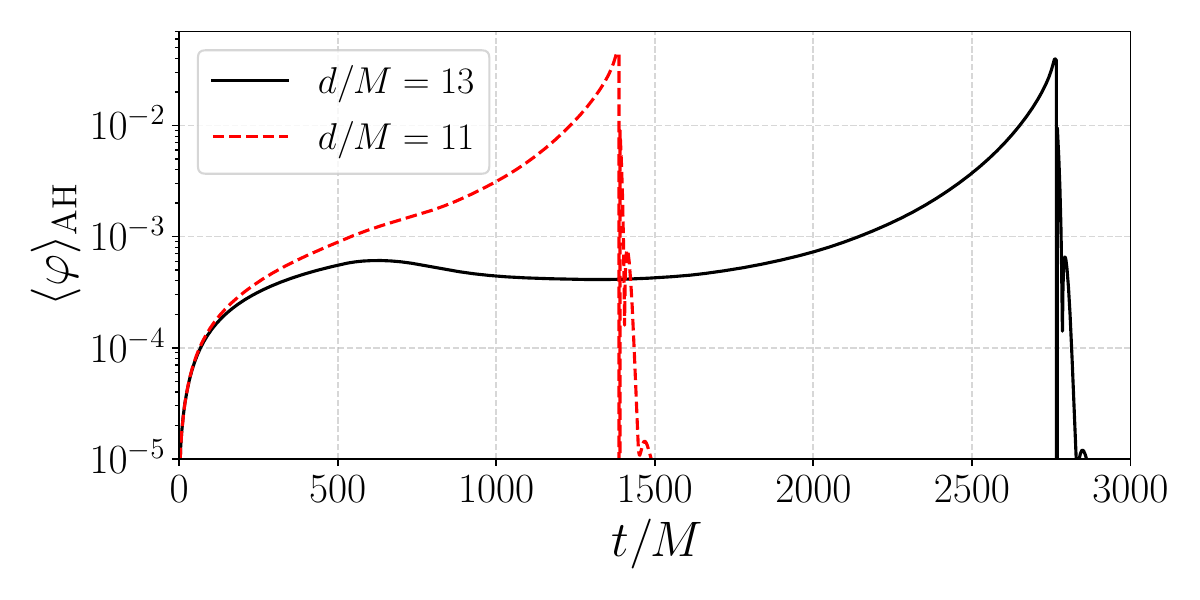}
\caption{The scalar field development for $(\lambda/M)^2=0.688$, $\beta=48$ with initial data ID1, i.e. a uniform initial scalar field with an amplitude of $10^{-5}$ throughout all the space. The initial puncture distance is different for the two simulations, namely $d/M=11$ and $d/M=13$. Both simulations stop shortly after the merger when the scalar field drops to zero because the newly formed BH is too massive to sustain scalar field hair. (top) The average value of the scalar field at the apparent horizon (AH) of one of the two BHs as a function of the distance between the BHs throughout inspiral. When the merger occurs (which happens at $t\sim 1350\,M$), we compute it instead at the horizon of the remnant. (bottom) The scalar field as a function of simulation time on a logarithmic scale. }
\label{fig:SF_Develop}
\end{figure}

Having discussed the simulation setup, let us focus now on how DS develops in actual NR merger simulations. What we have observed in  Fig.~\ref{fig:DS_different_lambda2}, based on semi-analytic calculations, is a nearly instantaneous development of the scalar field once the DS radius for these specific values of the parameters is approached. These calculations assume, however, a quasi-adiabatic evolution. In actual numerical simulations, it turns out that the scalar field growth time can be larger than the time to merger. 

With the available computational resources and speed of the code, we are able to simulate a full BBH merger starting from a maximum separation of roughly $15\,M$. Therefore, as a first setup, we have considered the first and second BBH simulations in Table \ref{tab:sim_params}. They are obtained for $(\lambda/M)^2=0.688$, $\beta=48$ and initial data ID1, which constitutes a constant scalar field of $10^{-5}$. The threshold for spontaneous scalarization, i.e. scalarization of isolated BHs,  is $(\lambda_{DS}/M)^2=0.7255$ \cite{Doneva:2017bvd,Blazquez-Salcedo:2018jnn,Silva:2017uqg} and, as $(\lambda/M)^2$ approaches this value, the DS radius tends to infinity. Thus, $(\lambda/M)^2=0.688$ is relatively far away from this limit, and the threshold for DS lies at $d=11.71\,M$. 
\begin{figure}
\centering
\includegraphics[width =0.48\textwidth]{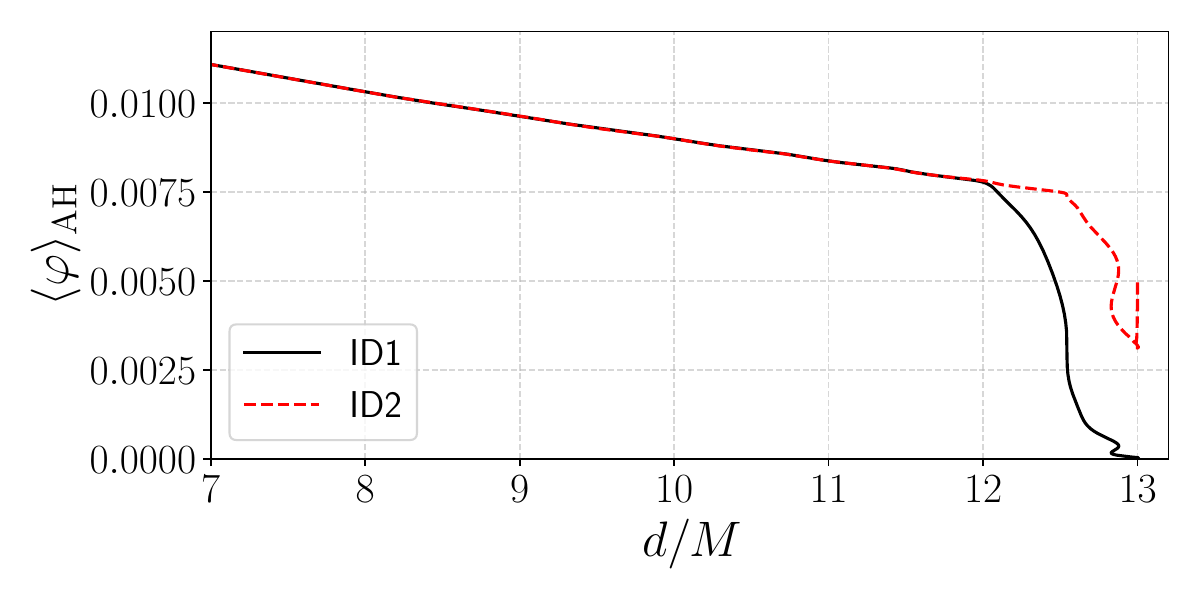}
\includegraphics[width =0.48\textwidth]{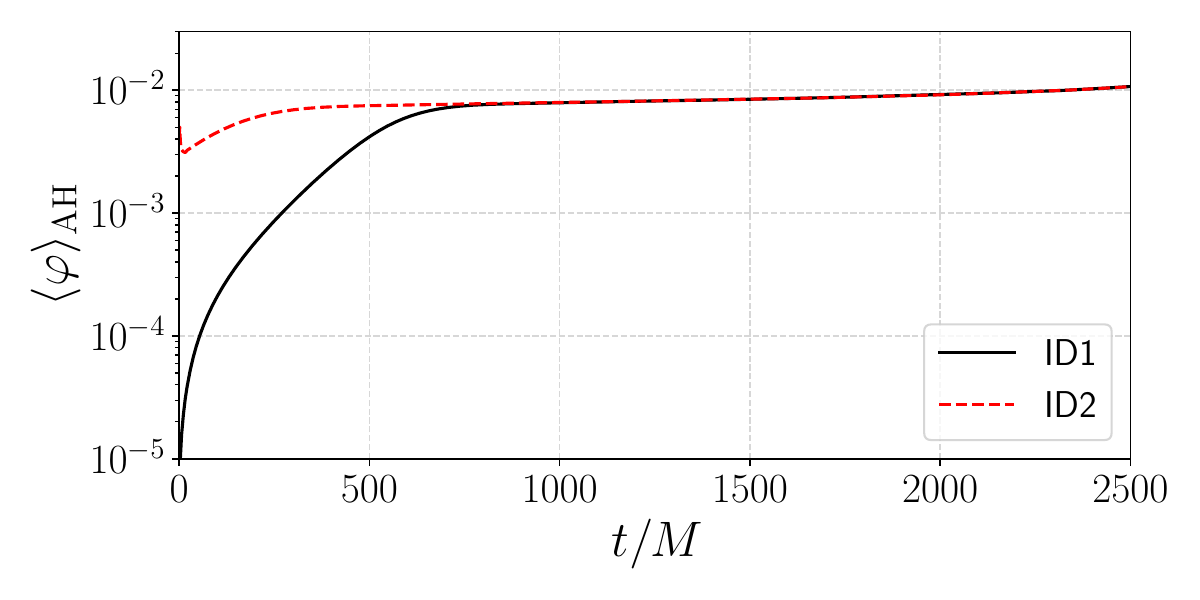}
\includegraphics[width =0.48\textwidth]{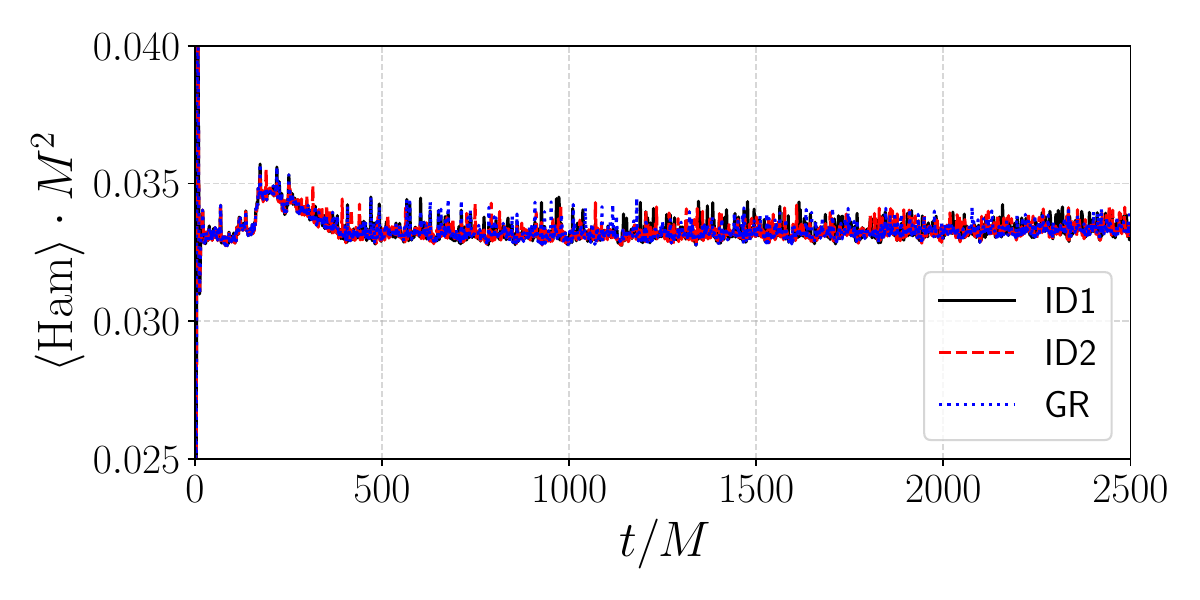}
\caption{The scalar field development for $(\lambda/M)^2=0.712$, $\beta=800$, and the two types of initial data described before. 
(top) The average value of the scalar field at the apparent horizon as a function of the distance between the BHs. (middle) The scalar field as a function of simulation time. (bottom) The average value (with the $L^2$ norm) of the Hamiltonian constraint throughout the whole space.}
\label{fig:ComparePhiID}
\end{figure}
Ideally, one should start the inspiral from a larger distance with a small scalar field perturbation and let the system evolve. As it turns out, though, the growth time of the scalar field is too long, so the evolution is not adiabatic. This is demonstrated in Fig.~\ref{fig:SF_Develop} (the simulation with $d/M=13$, black line). Clearly, the seed scalar field decreases (after a first increasing phase) until the orbital separation reaches a threshold distance of roughly $d=11.71\,M$, corresponding to the threshold for DS. After that, it starts to increase nearly exponentially as a function of time. This exponential growth should last until the scalar field value has increased enough that nonlinear  mechanisms damping the scalarization set in, saturating to a nearly constant value. This is not the case here, though, since the growth is too slow and the scalar field does not have enough time to fully develop. Instead, the scalar field starts  increasing even more rapidly as the two BHs approach, and it never reaches quasi-equilibrium. The faster increase is explained by the fact that the growth time of the scalar field gets shorter as the distance between the BHs decreases.  

This picture repeats if we start from a smaller distance, e.g. $d/M=11$ (red line in  Fig.~\ref{fig:SF_Develop}). Here, instead, the scalar field starts increasing right away, because at this distance  DS is already operating. However, again, it does not reach quasi-equilibrium. This is the reason why at merger the scalar field is different in the two simulations -- since no quasi-equilibrium was reached, the final value  depends on the initial data. These numerical results show that the semi-analytic approach developed in the previous section should be interpreted as an approximation, because it does not take into account the finite speed of the scalar field growth. Thus, it is reliable only for setups where the DS radius is large, and the scalar field has time to fully develop before merger.

As a next step, let us discuss a model with a larger coupling strength, namely $(\lambda/M)^2=0.712$ and $\beta=800$ (the $3^{\rm rd}$ and $4^{\rm th}$ models in Table \ref{tab:sim_params}). The motivation behind this parameter choice is the following: $(\lambda/M)^2$ is closer to the threshold of $(\lambda_{\rm DS}/M)^2=0.7255$ and, thus, DS occurs at a larger distance $d/M=33.25$, with shorter growth time. The high value of $\beta=800$ is motivated by the fact that, for the coupling function \eqref{eq:coupling}, higher $\beta$ leads to smaller scalar field equilibrium values, which are faster to reach during numerical evolution.

\begin{figure}[t]
          \centering
\includegraphics[width =0.48\textwidth]{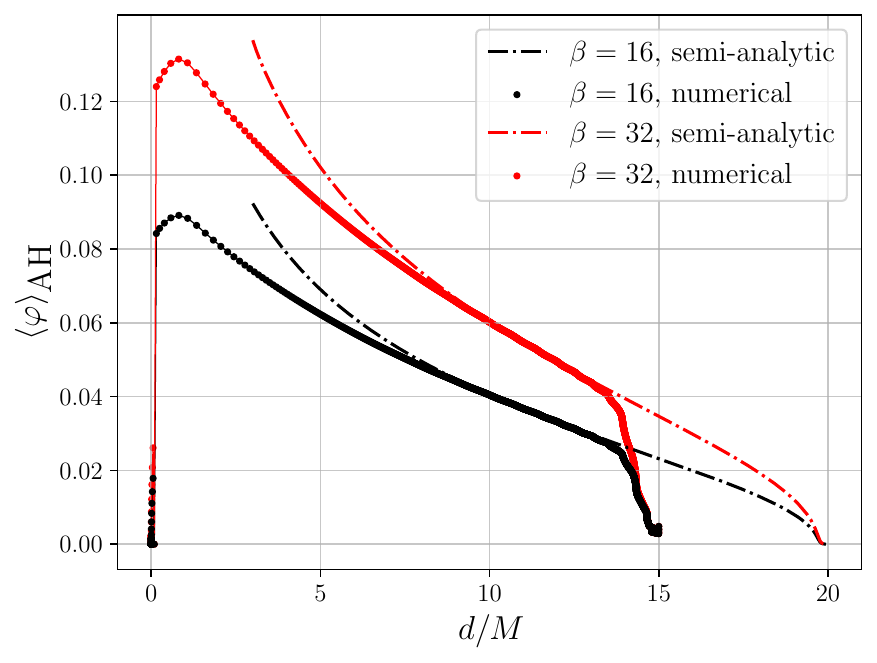}
 \caption{The average value of the scalar field at the apparent horizon as a function of distance for a BBH merger with $(\lambda/M)^2=0.703$ and $\beta=16$ (black dots) and $\beta=32$ (red dots). The numerical evolution has been started at $d/M=15$ due to the high computational costs, and ID2 is used (a Gaussian scalar field pulse around the two BHs) in order to shorten the time until the scalar field has fully developed. We compare these simulations with the scalar field at the event horizon in the semi-analytic approach for $\beta = 16$ (black dot-dashed line) and $\beta = 32$ (red dot-dashed line). Notice that the event horizon and the apparent horizon coincide in the static limit (implicitly assumed by the semi-analytic approach). }
\label{fig:Comparison}
\end{figure}

In Fig.~\ref{fig:ComparePhiID}, we present the scalar field evolution starting from $d/M=13$. Clearly, this is smaller than the threshold DS separation listed in Table \ref{tab:sim_params}. This is done purely for practical purposes (due to limited computational resources). Nevertheless, as we will discuss below, one is able to obtain a reliable scalar field evolution. The middle panel of the figure demonstrates that for both types of initial data, the scalar field starts growing exponentially after the initial settling of the system, because the chosen distance is shorter than the DS threshold. At a certain point, the growth is quenched, and the scalar field saturates to a quasi-equilibrium value that increases slowly as the distance between the two BHs decreases. Even though the path to reach this quasi-equilibrium value is very different for the two initial data, the scalar fields, once fully developed, agree perfectly. 
\begin{figure}[t]
\centering
\includegraphics[width =0.48\textwidth]{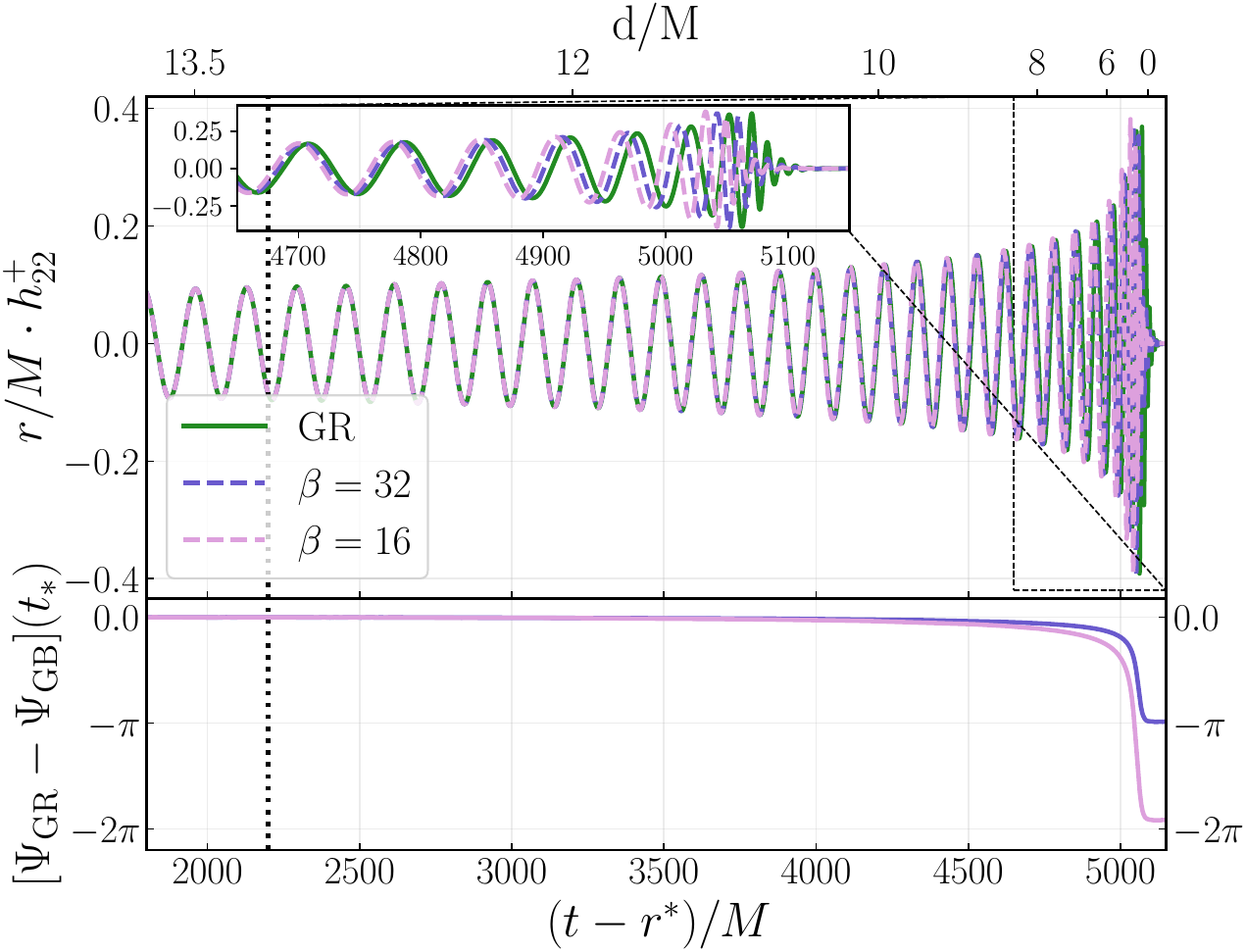}
\caption{The $l=2, m=2$ mode of the  plus polarization for GR, and for $(\lambda/M)^2=0.703$ (with $\beta=16$ and $\beta=32$), as functions of the retarded time $u=t-r^*$ (lower $x-$scale), where $r^*$ is the tortoise coordinate, and the distance between BHs in the GR simulation (upper $x-$scale). The strain is extracted at a distance of $140\,M$. The bottom panel shows the dephasing of the EsGB waveforms with respect to GR in the time domain.}
\label{fig:strain}
\end{figure}
The lower panel of Fig.~\ref{fig:ComparePhiID} also shows the Hamiltonian constraint, averaged over the whole computational domain, as a function of time. In addition to the two EsGB simulations with different initial data, a GR simulation with zero scalar field is also plotted. As  can be seen, even though the initial scalar field profile is not constraint-satisfying, the Hamiltonian constraint violation is practically the same in the three cases. This proves that, indeed,  constraints damping is efficient, and the results are reliable.

We conclude this subsection with a comparison between the semi-analytic approach and the NR simulations. This is an important consistency check, and provides an idea of how much one can trust the semi-analytic predictions, especially at very large separations where we cannot run full NR simulations. While at smaller separations some mismatch between the two predictions is expected, due to the approximations that the semi-analytic computation requires, it is important to check that the predicted value of the scalar field right after DS is robust. In fact, due to the very high computational cost of the NR merger simulations, in the next section, where we will assess the detectability of DS, we will  rely on the semi-analytic calculation. In  Fig.~\ref{fig:Comparison} we show the scalar field development in the two cases, for $(\lambda/M)^2=0.703$ and $\beta = 16$ (black dots) and $\beta = 32$ (red dots),  corresponding to the $5^{\rm th}$ and $6^{\rm th}$ simulations in Table~\ref{tab:sim_params}.
It can be clearly observed that when the simulation starts at a separation smaller than the scalarization radius predicted semi-analytically, the scalar field at the horizon quickly settles to a quasi-equilibrium value and starts growing slowly.\footnote{The scalar field settles to a quasi-equilibrium more easily in this case also because the simulations start from a larger separation. Thus, the distance between the BHs decreases much slower at $d/M=15$ compared to $d/M=13$ for example, and the scalar field has more time to develop.} This growth agrees well with the one predicted semi-analytically, represented by the black ($\beta=16$) and red ($\beta=32$) dot-dashed curves, up to $d\simeq 6 \,M$, where the adiabatic approximation begins to break down. This is also consistent with purely PN considerations, since the next-to-leading-order correction in a $M/d$ expansion starts being increasingly relevant at $d \sim\mathcal{O}(5-10)\,M$. Indeed, for very small separations, the semi-analytic computation tends to systematically overestimate the scalar field growth. Moreover, the BBH separation is not a gauge-invariant quantity and thus the ones extracted from the simulations are expected to deviate as we get closer to the plunge. On the other hand, this difference tends to be negligible for $d\gtrsim 6M$.

\begin{figure*}[t]
\centering
\includegraphics[width =0.48\textwidth]{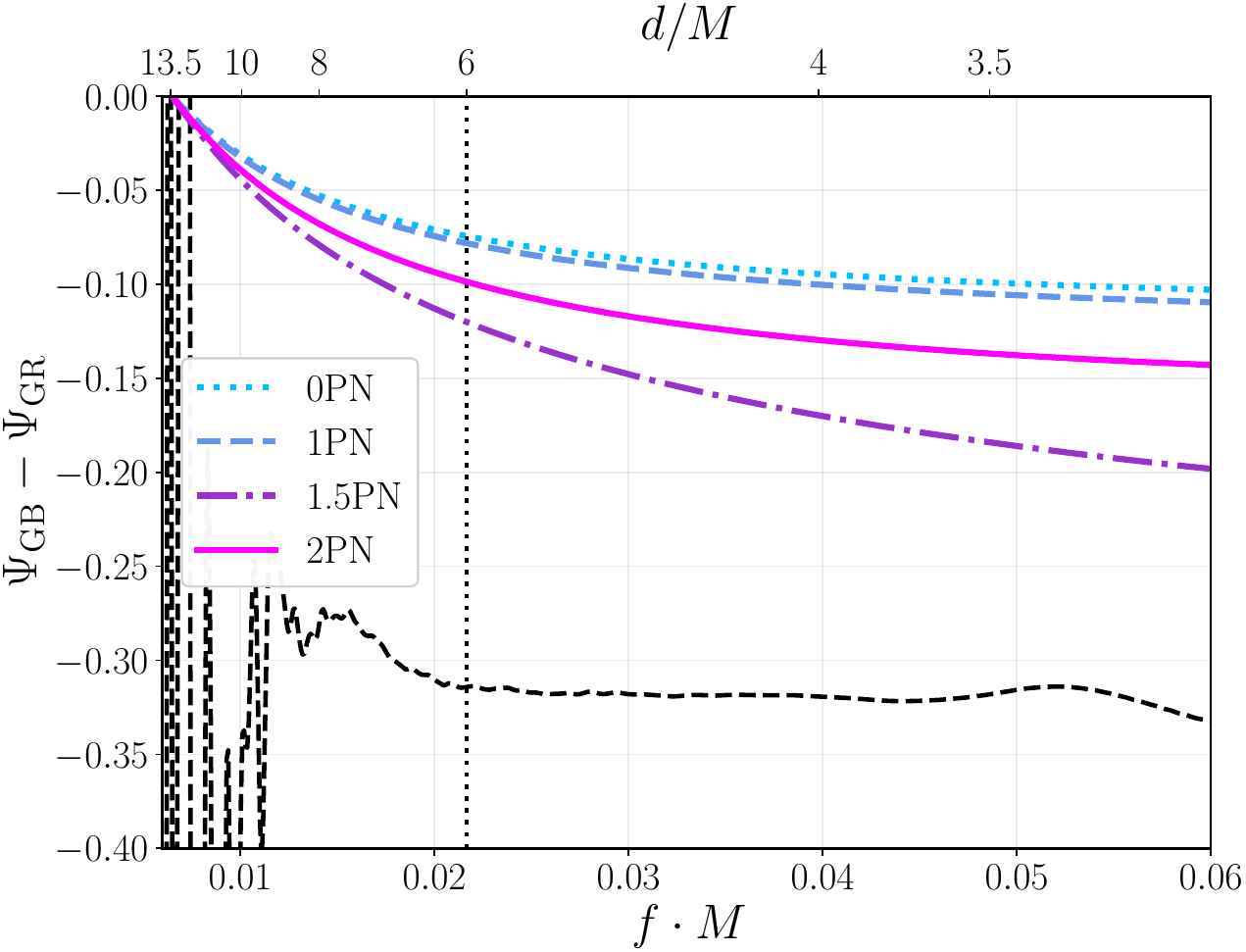}
\includegraphics[width =0.48\textwidth]{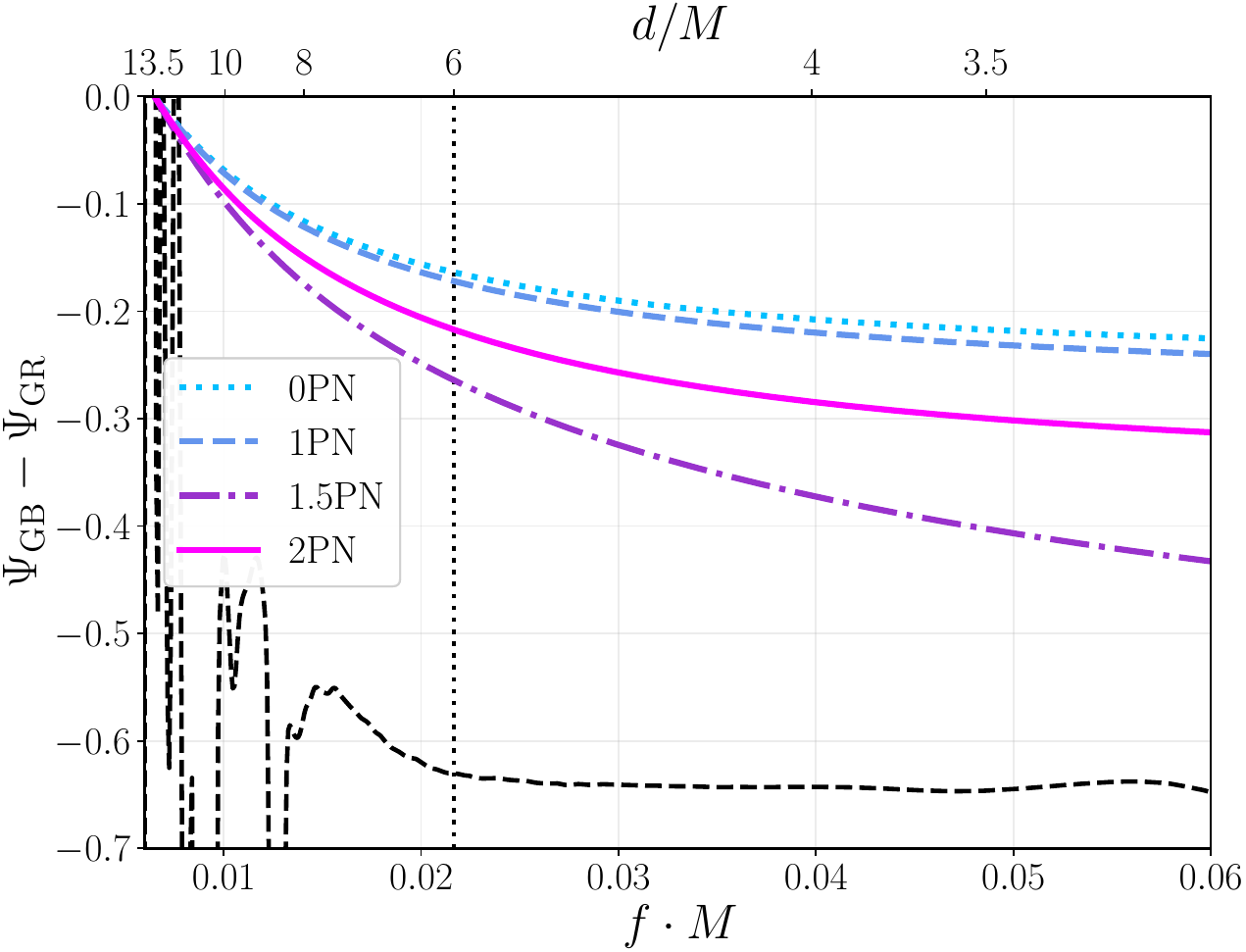}
\caption{Dephasing between GR and EsGB gravity in frequency domain for the simulations presented in Fig.~\ref{fig:strain} ($\beta=32$ on the left and $\beta=16$ on the right). We also show the PN prediction for the dephasing calculated at different PN orders. Our results show an agreement in order of magnitude between the PN dephasing and our non-linear simulations. The top axis shows the distance between BHs corresponding to every frequency. The vertical dotted line marks a distance of $d/M=6$ roughly below which the PN predictions are not reliable, according to the results in Fig.~\ref{fig:Comparison}.  }
\label{fig:dephasing}
\end{figure*}

We conclude that, at large  separations and especially for $d>10 \,M$, the scalar field predicted by the semi-analytic approach is in  very good agreement with NR simulations. While a full inspiral-merger simulation can often be performed only starting from a  distance smaller than $d_{\rm DS}/M$, due to the high computational costs, our tests show that we can safely trust the scalarization radius predicted semi-analytically. We conclude that the semi-analytic model can be a useful and computationally cheap tool for exploring the behavior of the scalar field at very large separations.


\section{Gravitational wave strain and dephasing with respect to GR}
\label{sec4}

In this section, we present the GW strain simulated for 
 $(\lambda/M)^2=0.703$, $\beta=32$ and $d/M=15$ (corresponding to the $5^{\rm th}$ case in Table \ref{tab:sim_params}), for which the scalar field develops fast and reaches quasi-equilibrium. We also compare it with the strains obtained with $\beta=16$ and with $\lambda=0$ (i.e., GR), corresponding to the $6^{\rm th}$ and $7^{\rm th}$ cases of Table \ref{tab:sim_params}, respectively. The comparison is shown in Fig.~ \ref{fig:strain}, where one can see a small but observable dephasing of the EsGB waveforms from GR.  
 As expected, the phase difference  is larger for  lower $\beta$. The signals have been aligned in frequency and phase over a time window ranging from $t_i=1800M$ to $t_f=2200M$ (the latter shown by the vertical dotted line in the plot), as done in~\cite{Julie:2024fwy,Corman:2025wun}. This time window was chosen in such a way that the scalar field has fully developed, i.e. the system had reached quasi-equilibrium, while still one can observe around 13-14 cycles before merger. The latter value, $t_f=2200M$, is shown by the vertical dotted line in the plot.  We note that the sign of the dephasing coincides with the same as expected from PN calculations, in contrast to what was recently found in \cite{Corman:2025wun}, even after performing the alignment in the same way as in \cite{Corman:2025wun}. We attribute this to both a weaker coupling and to the different dynamics induced by the change of the scalar charge throughout the inspiral.

Next, we measure the dephasing  as a function of frequency, comparing the same initial data with zero and nonzero coupling $\lambda$. We also compare this with PN calculations, as it is done in~\cite{Corman:2022xqg,AresteSalo:2025sxc}. As explained more thoroughly in~\cite{AresteSalo:2025sxc}, we express the orbital phase $\Psi$, defined as half the complex phase of the (2,2) mode of the Weyl scalar, as a function of the frequency of the GW. 

Then we compare those values with the PN expressions for the orbital phase dephasing up to 2PN order, which were also used in~\cite{AresteSalo:2025sxc} and which can be found in Appendix B of~\cite{Corman:2022xqg}, with the following modifications. Let us denote the PN expressions for the GW dephasing in the shift-symmetric case of EsGB gravity (i.e., when $f(\varphi)=\varphi$) by
\begin{eqnarray}
    \Delta\Psi_{\rm ss}(f)=Q_{\rm ss}^2\sum\limits_ic_i (f^{(-5+2i)/3}-f_0^{(-5+2i)/3})\,,
\end{eqnarray}
where $c_i$ are some given PN coefficients,\footnote{Note that the definition of $c_i$ is slightly different from \cite{Corman:2022xqg} because of the different normalizations we use.} all of them independent of the frequency and $Q_{\rm ss}$ is the shift symmetric theory scalar charge. It is a assumed to be a constant during evolution and, in the perturbative with respect to $\lambda$ regime, it has an analytic expression proportional to the GB coupling $\lambda^2$. In the DS case, the scalar charge is no longer a constant but instead it can increase significantly prior to the merger. Therefore, we can write an approximate expression for the GW dephasing for the DS of the form
\begin{eqnarray}
    \Delta\Psi(f)=\int_{f_0}^fQ^2(f')\sum\limits_i\frac{-5+2i}{3f'} c_if'^{(-5+2i)/3} df'\,.
\end{eqnarray}
Here $Q(f)$ is the scalar charge for the DS case studied in this paper, for which we use the results from the semi-analytic computation presented in Sec.~\ref{sec2}.


In Fig.~\ref{fig:dephasing}, we show, for both $\beta=32$ and $\beta=16$, the comparison between GW dephasing computed with the PN approach and the numerical one. The comparison starts at a time where the scalar field has fully developed and the two NR waveforms in each panel, the GR one and the scalarized one, are properly aligned following \cite{Corman:2025wun} as we discussed above. Clearly, the total dephasing would be bigger if we were starting from the DS radius, which is at $d/M=19.75$, corresponding to a frequency of roughly $f_0M=0.0036$. In additon, one can notice that the NR dephasing looks very ``noisy''. Clearly, this is not the case in the lower panel of Fig. \ref{fig:strain} where the dephasing is plotted as a function of time. The somewhat large variations here are caused by inaccuracies when extracting the GW frequency from the numerical data especially for small coupling parameters and thus relatively small dephasing.

As one can see in both panels, the PN dephasing differs by a factor of $\sim2-3$ from the dephasing extracted from the NR simulation with the same parameters. This mismatch is similar to one obtained, for instance, in the neutron star DS studies~\cite{Sampson_2014}. While it is important to be aware of this systematic bias, the PN approach can still be useful for order-of-magnitude estimates, especially in cases in which a wide exploration of parameter space is pursued. Indeed, in the next section, we are going to employ the PN predictions for estimating the observability of DS at different BBH masses, since running a full NR simulation for every mass bin would have had a prohibitive computational cost.

\begin{figure}[t]
          \centering
\includegraphics[width =0.48\textwidth]{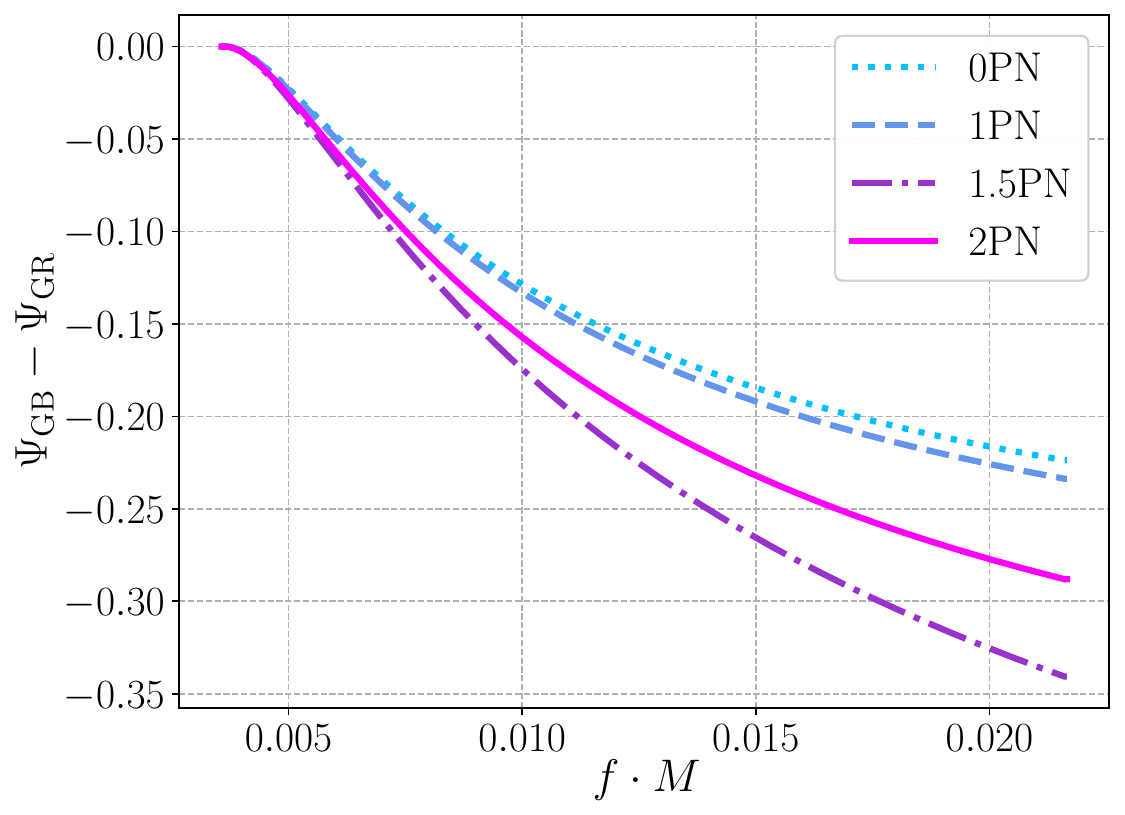}
 \caption{Dephasing between GR and EsGB gravity in frequency domain for $(\lambda/M)^2=0.703$, and $\beta=16$, at different PN orders. }
\label{fig:dephasing_DS}
\end{figure}

In Fig.~\ref{fig:dephasing_DS}, we show the dephasing at different PN orders, for the same parameters $\beta = 16$ and $(\lambda/M)^2=0.703$ as in Fig.~\ref{fig:dephasing}, but starting from a larger distance, namely the scalarization radius $d_{\rm DS}$. In that case, the NR simulation results are missing since the computational cost for calculating a merger starting from $d_{\rm DS}$ is too high. We see that the total phase difference is about twice larger compared to the shorter inspiral in Fig.~\ref{fig:dephasing}. Therefore, even though the dephasing accumulates very slowly at large separation, the number of orbits increases significantly as we increase the distance between the BHs and eventually leads to a non-negligible phase difference. 



\section{Detectability}
\label{sec5}

A computationally cheap estimate of the detectability of DS can be conveniently performed in the frequency domain. The Fourier transform of the GW strain is defined as
\begin{equation}
    \tilde{h}(f)=\int_{-\infty}^{+\infty} {\rm d}t\,e^{-2\pi i ft}h(t)\,.
\end{equation}
Notice that the GW frequency $f$ is approximately twice the orbital frequency, at least at the early inspiral stage.
The inspiral part of the GW template in the frequency domain can be computed analytically by employing the stationary phase approximation, assuming that the phase varies much faster than the amplitude of the wave. With this approach, one obtains in GR~\cite{Cutler:1994ys}
\begin{equation}
    \tilde{h}(f)=A(f) e^{i \Psi(f)}\,.
\end{equation}
The wave amplitude reads
\begin{equation}
    A(f) = -\left(\frac{5}{384}\right)^{1/2}\frac{\pi^{-2/3}\mathcal{M}_c^{5/6}}{d_L}\mathcal{F}f^{-7/6}\,,
    \label{eq:GW_amplitude}
\end{equation}
where we introduced the chirp mass $\mathcal{M}_c\equiv M \eta^{3/5}$, with $\eta=m_1m_2/(m_1+m_2)$ being the symmetric mass ratio, and the luminosity distance $d_L$. The geometrical factor $\mathcal{F}$ accounts for the detector response and the system inclination with respect to the line of sight. The GW phase can be estimated in the stationary phase approximation as~\cite{Cutler:1994ys,Yunes_2009}
\begin{equation}
    \Psi(f)=2\pi\int^f{\rm d}f'\,\left(1-\frac{f}{f'}\right)\frac{f'}{\dot{f}(f')}\,.
\end{equation}
The time evolution of the GW frequency $\dot{f}$ can be computed analytically at a given PN order, and it is strongly related to the leading channel of energy radiation. 

Due to the presence of scalar charges,  BBHs can radiate energy through both tensor and scalar waves. The leading order term in the scalar flux is the dipole, which is proportional to the squared difference of the scalar charges of the two objects. For this reason, if a BBH system exhibits a nonvanishing difference of scalar charges at large  separations, this contribution generally dominates, as it enters at lower PN order, and is therefore less suppressed with the separation. In particular, comparing the power emitted through scalar  ($\dot{E}_{\rm scal}$) and tensor  ($\dot{E}_{\rm GW}$) waves at leading PN order, one finds that the scalar dipole dominates when~\cite{Shiralilou:2021mfl}
\begin{equation}
    \Delta\alpha^2\gtrsim\frac{96}{5}\frac{G_{\rm eff}M}{d}\,,
    \label{eq:dipole_domination}
\end{equation}
where we defined the effective gravitational coupling $G_{\text{eff}}\equiv 1+\alpha^2$, with $\alpha$ being the dimensionless scalar charge (see Appendix~\ref{Comparison_Juliè}), and $\Delta\alpha = \alpha_1-\alpha_2$.
However, in the case of DS, as discussed in the previous section, the scalar charges are only excited at separations equal or smaller than the scalarization radius. At the onset of DS, these charges are initially very small, 
and Eq.~\eqref{eq:dipole_domination} is therefore not satisfied. If one moves closer to the plunge, 
where scalar charges are larger, 
the separation on the right-hand side of Eq.~\eqref{eq:dipole_domination} is smaller, and 
Eq.~\eqref{eq:dipole_domination} cannot be satisfied either.

We therefore do not expect that the scalar dipole dominates,
 since either the left-hand side of Eq.~\eqref{eq:dipole_domination} is vanishing or very small (at large separations), or the right-hand side is very large (at small separations). In other words, unlike the case of systems with component masses $ < 0.587 \lambda$ (corresponding to a total mass such that $\lambda/M>(\lambda/M)_{\rm crit}$), for which the scalar charges are already present in the early inspiral phase, the systems that we are interested in are practically never dominated by dipole radiation. Hence, if we do not consider BBHs in which at least one of the two is already scalarized in isolation, the maximal deviation from GR is obtained in an equal mass system with $\lambda/M\simeq(\lambda/M)_{\rm crit}$, i.e., close to the limiting value of the coupling for which one can have DS. This is evident from  Fig.~\ref{fig:q098}, where we show that even a  tiny mass asymmetry ($q=0.98$) drastically decreases the scalarization radius. In fact, although the system with $q = 0.98$ and $(\lambda/M)^2=0.703$  produces a small amount of scalar dipole radiation, the DS effect is much more visible for the equal-mass system with $(\lambda/M)^2=0.717$. This further motivates our choice of performing NR simulations for mass-symmetric BBHs. 
From now on, we will continue to restrict our analysis to equal mass systems, i.e. we will fix $\eta=1/4$. 

We now want to quantify the observability of the dephasing produced by DS in comparison to a pure GR waveform. We will assume that such dephasing is the only observable signature of DS, while the amplitude of the GW is always given by Eq.~\eqref{eq:GW_amplitude}, derived within GR. Furthermore, since we want to set an upper bound for the observability of DS, we will consider an optimal geometrical configuration for which $\mathcal{F}\simeq 1$. For a given detector, characterized by a noise power-spectral density $S_n(f)$, the impact of the GW dephasing can be  quantified by the \emph{effective number of cycles}, defined as~\cite{Sampson_2014}
\begin{equation}
    \mathcal{N}_e\equiv \min\limits_{\delta t, \delta\phi} \left\{\frac{1}{2\pi \,{\rm SNR}}\left(\int_{f_{\rm min}}^{f_{\rm max}} {\rm d} f \,\frac{A(f)^2\Delta\Phi^2}{S_n(f)}\right)^{1/2}\right\}\,,
\end{equation}
with $\Delta\Phi = \Delta\Psi+2\pi f\delta t-\delta\phi$. The minimization over the time and phase shifts $\delta t$ and $\delta\phi$ can be carried out analytically. We also define the signal-to-noise ratio
\begin{equation}
    {\rm SNR}\equiv\left(\int_{f_{\rm min}}^{f_{\rm max}} {\rm d}f\, \frac{A(f)^2}{S_n(f)}\right)^{1/2}\,.
\end{equation}
Let us now introduce the GW frequency corresponding to the onset of DS, $f_{\rm DS}$, where a nonzero dephasing with respect to GR starts  accumulating,  and the GW frequency corresponding to the end of the inspiral, $f_{\rm end}$, 
i.e. where the PN approximation starts being unreliable for our choice of parameters. Then, if the detector that we are considering is sensitive to a frequency window $f \in (f_1,f_2)$, the integration limits will be respectively
\begin{equation}
\begin{split}
    &f_{\rm min}= {\rm max}\left\{f_1,f_{\rm DS}\right\} \,,\\
    &f_{\rm max}={\rm min}\left\{f_{\rm end},f_2\right\}\,.
\end{split}
\end{equation}
The orbital frequency corresponding to a given separation during the inspiral can be approximately computed through  Kepler's law
\begin{equation}
    f_{\rm orb}=\frac{1}{2\pi}\sqrt{\frac{M}{d^3}}\,.
\end{equation}
As a first-order approximation, one can consider the effect to be observable if the following criterion is fulfilled~\cite{Sampson_2014,Mangiagli:2018kpu},
\begin{equation}
    \mathcal{N}_e \gtrsim \frac{1}{\sqrt{2}\pi \,{\rm SNR}}\,.
    \label{eq:observability_criterion}
\end{equation}
In order to understand the meaning of this condition, let us recall the definition of Bayes factor (BF) between  two models (in our case GR and EsGB)
\begin{equation}
    {\rm BF}\equiv\frac{p\left(\mathcal{D}|\rm EsGB\right)}{p\left(\mathcal{D}|\rm GR\right)}\,,
\end{equation}
where $p$ indicates the probability of having a set of data $\mathcal{D}$, given a model.
The effective number of cycles, in the limit of small deviations from GR, can be related to the BF as~\cite{Sampson_2014}
\begin{equation}
    \ln({\rm BF})\sim 2\pi^2\,{\rm SNR}^2\,\min\limits_{\theta^i}\left\{\mathcal{N}_e^2\right\}
\end{equation}
where minimization is performed over the parameters of the GW event $\theta^i$. In practice, we only carry out the minimization over $\delta t$ and $\delta \phi $, which appear in the definition of $\mathcal{N}_e$, avoiding the one over the parameters $\theta^i$, which would be computationally much more expensive. Thus, our estimate for the BF has to be interpreted as an upper limit. A more detailed analysis, also accounting for the minimization over the system parameters, is beyond the scope of this work. Furthermore, we expect the minimization over those parameters to be largely irrelevant for our computation. In fact, the  dephasing between the GR and the EsGB waveforms due to DS accumulates at $d<d_{\rm DS}$. Therefore, shifting e.g. the masses to offset
that phase mismatch would produce a
 large difference in the early inspiral.

In conclusion, the criterion of Eq.~\eqref{eq:observability_criterion} is approximately equivalent to requiring that $\ln ({\rm BF}) \gtrsim1$, which means requiring  evidence in favor of EsGB with respect to GR~\cite{Jeffreys:1939xee,Kass:1995loi}.
Moreover, Eq.~\eqref{eq:observability_criterion} can also be interpreted in terms of \emph{faithfulness} of the GW templates in describing the signal (see e.g.~\cite{Mangiagli:2018kpu}).

As already anticipated, in the EFT perspective, we are likely to get the best constraints on the theory at smaller scales, namely from the late inspirals of stellar mass BHs, rather than at supermassive scales. For this reason, the most promising observations of DS are GW detectors targeting stellar-mass BBHs. So far, LIGO-Virgo observations have found no evidence for deviations from GR~\cite{LIGOScientific:2021sio,Wang:2024erh,KAGRA:2025oiz,LIGOScientific:2025obp}. However, future LIGO-Virgo-KAGRA (LVK) detectors, as well as future third-generation ground-based detectors, will provide more accurate constraints through inspiral-based tests of gravity. As a representative example, we will focus on the future ground-based detector Einstein Telescope (ET), which will target approximately the same mass-band as LVK, with a much higher sensitivity. 

As discussed above, the maximum observable effect is obtained for an equal-mass BBH system. In order to enhance the observable effect, we select a relatively small value of $\beta$, namely $\beta = 16$. In particular, fixing a value of the coupling $\lambda$, we divide the total mass domain into bins of $1\,M_\odot$, and consider BBHs with total mass such that $(\lambda/M)^2= 0.703$, which translates into $\lambda \sim 1.24\,{\rm km}\,(M/M_{\odot})$. This corresponds to the $6^{\rm th}$ simulation in Table \ref{tab:sim_params}, which is the case for which we compared the numerical and analytic dephasing in Fig.~\ref{fig:dephasing}. Assuming these conditions, we compute the maximum redshift $z_{\rm max}$ at which DS is potentially observable, according to the criterion of Eq.~\eqref{eq:observability_criterion}, and assuming a $\Lambda$CDM universe. Then, we repeat the procedure for the next mass bins, where the ratio $\lambda/M$ is smaller. This whole analysis is performed for five different values of the coupling, i.e. $\lambda = 24.75\,{\rm km}\,,\,37.13\,{\rm km}\,,\,49.50\,{\rm km}\,,\,61.88\,{\rm km}\,,\,74.25\,{\rm km}$. These values correspond to tuning the total mass of the BBH in such a way that the condition $(\lambda/M)^2= 0.703$ is verified, respectively, for $20\,M_\odot\,,\,30\,M_\odot\,,\,40\,M_\odot\,,\,50\,M_\odot\,,\,60\,M_\odot$. Note that, in order to maximize the deviation from GR, we have chosen the maximum considered mass to be as close as possible to the current observational limits for pulsars studied in \cite{Danchev:2021tew}. The latter yield slightly different results for  different equations of state, ranging from $\lambda\lesssim 21.79$ km to $\lambda\lesssim 50.87$ km. 
Nevertheless, slightly higher values of $\lambda$ are also allowed, as pointed out in \cite{Wong:2022wni}, where it was shown that strongly disfavored values start at $\lambda\gtrsim 82.55$ km.

\begin{figure}[t]
          \centering
\includegraphics[width =0.48\textwidth]{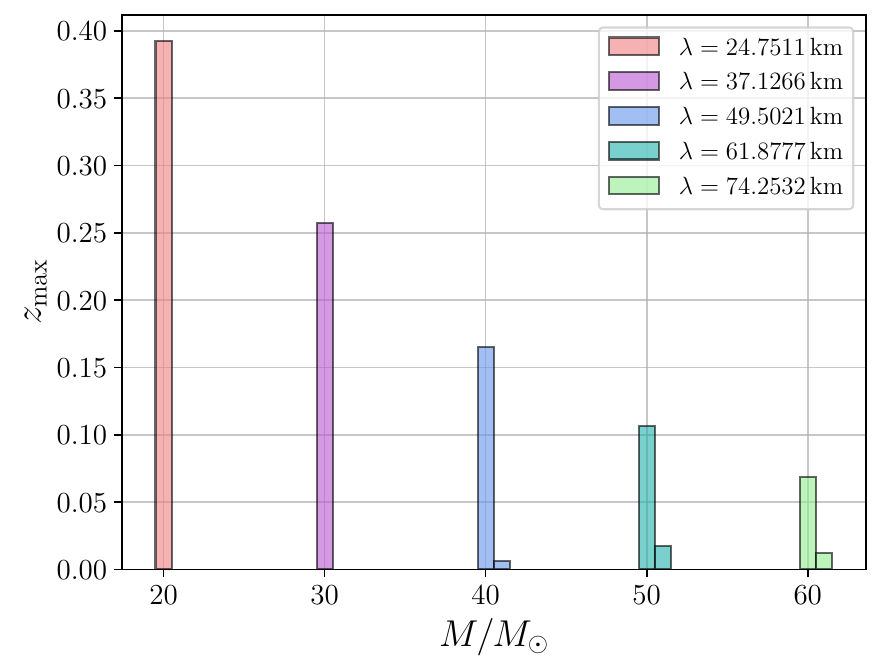}
 \caption{Maximum redshift for which DS is potentially detectable with ET, for $\beta=16$ and different values of $\lambda$, as a function of the total mass of the binary. The x axis is divided in bins of $1 \,M_\odot$. We considered four different values of $\lambda$, such that $\lambda^2/0.703=(20\,M_\odot)^2\,,\,(30\,M_\odot)^2\,,\,(40\,M_\odot)^2\,,\,(50\,M_\odot)^2\,,\,(60\,M_\odot)^2$. These values correspond, in units of length, to $\lambda = 24.75\,{\rm km}\,,\,37.13\,{\rm km}\,,\,49.50\,{\rm km}\,,\,61.88\,{\rm km}\,,\,74.25\,{\rm km}$. We compute the maximum redshift for the mass bin in which the effect is optimized, and then for successive mass bins. 
}
\label{fig:Detectability}
\end{figure}
Our results for detectability are shown in  Fig.~\ref{fig:Detectability}. It can be clearly observed that for every value of the coupling $\lambda$, DS is more likely to be observed in a mass range where the ratio $\lambda/M$ is close to the limiting value $(\lambda/M)_{\rm crit}$, as already discussed. Moreover, it can be observed that the mass window in which the effect is visible gets slightly wider as one moves to higher values of $\lambda$. On the other hand, the observability of the effect is also strongly affected by the sensitivity curve of the detector, and so at small masses, where the sensitivity of ET is peaked, the maximum redshift for which DS is theoretically observable is much larger. 

According to the predictions by LIGO-Virgo~\cite{KAGRA:2021duu,LIGOScientific:2025gwtc4}, we can expect a merger rate density of $\sim \mathcal{O}(1-10)\,{\rm Gpc}^{-3}\,{\rm yr}^{-1}\,M_{\odot}^{-1}$ for BBHs with component masses $\sim 10 \, M_{\odot}$. Assuming $\Lambda {\rm CDM}$ cosmology, the cosmic volume corresponding to $z\simeq 0.4$ is $\simeq15.64 \,{\rm Gpc}^3$. Therefore, we can expect a number of $\sim \mathcal{O}(10^1-10^2)$ merger events per year in this region. However, only a very small fraction of these events is going to provide a good candidate for observing DS. In fact, the estimated peak of the mass ratio is likely to be around $m_1/m_2\sim 0.75$, while only $\sim1\%-5 \%$ of the events are expected for $q>0.9$~\cite{LIGOScientific:2025gwtc4}.

We conclude this discussion by stressing again two aspects of the results presented in  Fig.~\ref{fig:Detectability}.
\begin{enumerate}
    \item Our estimate of $z_{\rm max}$ must be  viewed as an order of magnitude estimate. In realistic systems, the geometrical configuration is not optimal, and the covariances among different astrophysical parameters would further reduce the BF of EsGB with respect to GR. 

    On the other hand, as shown in Fig.~\ref{fig:dephasing}, the PN approach seems to underestimate the GW dephasing, resulting in a smaller predicted $z_{\rm max}$. These two different biases on the maximum redshift have opposite sign and are therefore competing.
    \item Fig.~\ref{fig:Detectability} does not represent the observability of BH hairs in EsGB in general, but only in the DS scenario. In other words, for different component masses and/or different mass ratios, a more promising observational strategy to test EsGB would be to look for dipole fluxes~\cite{Chamberlain:2017fjl}, due to either induced or spontaneous scalarization. In fact, as already discussed, dipole radiation, which is practically irrelevant for the considered BBH systems exhibiting DS, represents, in general, the dominant contribution in a PN expansion, and thus it is potentially the largest source of deviation from GR for other ranges of parameters.  
\end{enumerate}  

\section{Conclusions}

Within the family of ST theories of gravity, some specific models can evade classical BH no-hair theorems. The most prominent example is given by EsGB gravity. In this theory, deviations from GR are typically expected in GW signals already at the inspiral phase.

In this work, we focus on DS, i.e. the case in which initially uncharged BHs in a binary system acquire scalar charges as they reach a critical separation, which we refer to as scalarization radius. This phenomenon, already known in the context of neutron stars in the classical ST theories~\cite{Barausse_2013, Shibata:2013pra, Palenzuela_2014,Sampson_2014}, was first shown to occur for BBHs in EsGB in~\cite{Julie:2023ncq} for values of the coupling parameters close to, but still below, the threshold for the development of spontaneous scalarization of isolated BHs. 

We implement a simple and computationally inexpensive semi-analytic model, inspired by~\cite{Palenzuela_2014}, which allows us to estimate the behavior of the scalar charges of BBH as function of separation. The main assumption is that the system evolves adiabatically, meaning in this context that the total Wald entropy is constant. This is a fair assumption in the early inspiral. We also performed the first fully nonlinear NR simulations of BBH systems in EsGB gravity with DS using the recent advances described in \cite{AresteSalo:2025sxc}. Our simulations show very good agreement with our semi-analytic approach, for systems in which the scalarization radius is large enough, as shown in  Fig.~\ref{fig:Comparison}. It is interesting to note that, in contrast to previous results in shift-symmetric sGB gravity \cite{Corman:2025wun}, in this case we obtained a qualitative agreement between the PN and NR dephasing with respect to GR. Namely, the DS binaries merge faster. This can probably be attributed to the nature of the DS process, where the system starts from a true GR state and gradually develops a scalar field as the separation decreases. Another important difference with respect to \cite{Corman:2025wun} is the weaker scalar charge for the considered dynamically scalarized binaries.

When the scalarization radius predicted by the semi-analytic model is smaller than say $\lesssim 10-15\,M$, numerical simulations suggest that the scalar field does not have enough time to develop. Therefore,  DS does not reach a quasi-equilibrium before merger.
We interpret this mismatch between the two predictions -- numerical and semi-analytic -- as due to the breakdown of the adiabatic regime, which occurs in the last cycles of the BBH inspiral.



We also semi-analytically compute the dephasing between the GW signal in GR and EsGB in the frequency domain, generalizing the PN formulas reported in~\cite{Corman:2022xqg}, and finding results within the order of magnitude of the NR simulations. We finally employ the semi-analytically computed dephasing to estimate the observability of DS for different systems. We used a criterion based on the effective number of cycles of difference between GR and EsGB. This approach was inspired by~\cite{Sampson_2014}, and can be linked to the definition of Bayes factor. With a simple argument, we show that the most promising sources for observing DS are nearly-equal-mass BBHs. We conclude that DS will be potentially observable in this kind of sources with third-generation ground-based detectors such as ET, in a narrow mass window around the value corresponding to the scalarization threshold   of isolated BHs. 

\section{Acknowledgments}

We thank Rohit Chandramouli for insightful discussions on detectability criteria and waveform templates. We thank Maxence Corman for sharing the scripts to perform the alignment of the waveforms. We are also grateful to Félix-Louis Julié for useful discussions on the Wald entropy and for providing helpful comments on the manuscript. We thank the entire \texttt{GRTL} Collaboration\footnote{\texttt{www.grtlcollaboration.org}} for their support and code development work. This study is partly financed by the European Union-NextGenerationEU, through the National Recovery and Resilience Plan of the Republic of Bulgaria, project No. BG-RRP-2.004-0008-C01. The work by LC has been partially supported by the MUR FIS2 Advanced Grant ET-NOW (CUP: B53C25001080001) and by
the INFN TEONGRAV initiative. LAS is partly funded by Interuniversitaire Bijzonder Onderzoeksfonds (IBOF)/21/084. DD acknowledges financial support via an Emmy Noether Research Group funded by the German Research Foundation (DFG) under grant no. DO 1771/1-1, the Spanish Ministry of Science and Innovation via the Ram\'on y Cajal programme (grant RYC2023-042559-I), funded by
MCIN/AEI/10.13039/501100011033, and the Spanish Agencia Estatal de Investigaci\'on (grant PID2024-159689NB-C21) funded by MICIU/AEI/10.13039/501100011033 and by FEDER / EU. EB acknowledges support from the European Union’s Horizon ERC Synergy Grant ``Making Sense of the Unexpected in the Gravitational-Wave Sky'' (Grant No. GWSky-101167314) and the PRIN 2022 grant ``GUVIRP - Gravity tests in the UltraViolet and InfraRed with Pulsar timing''. We acknowledge Discoverer PetaSC and EuroHPC JU for awarding this project access to Discoverer supercomputer resources. This work used the DiRAC@Durham facility managed by the Institute for Computational Cosmology on behalf of the STFC DiRAC HPC Facility (www.dirac.ac.uk). The equipment was funded by BEIS capital funding via STFC capital grants ST/P002293/1, ST/R002371/1 and ST/S002502/1, Durham University and STFC operations grant ST/R000832/1. DiRAC is part of the National e-Infrastructure. We also used the resources and services provided by the VSC (Flemish Supercomputer Center), funded by the Research Foundation - Flanders (FWO) and the Flemish Government.

\appendix
\section{Equations of motion for static spherically symmetric black holes}
\label{Appendix:EOM}
The variation of the action~\eqref{eq:action} with respect to the metric tensor and the scalar field, gives respectively
\begin{equation}
\begin{split}
    &G_{\mu\nu}+ \Gamma_{\mu\nu} = 2\nabla_{\mu}\varphi\,\nabla_{\nu}\varphi
   - g_{\mu\nu}\,\nabla_{\alpha}\varphi\,\nabla^{\alpha}\varphi,\\
&\nabla_{\alpha}\nabla^{\alpha}\varphi = -\tfrac{1}{4}\lambda^{2}\,f'(\varphi)\,
   \mathcal{G}\,,
\end{split}
\end{equation}
where $\nabla_{\mu}$ denotes the covariant derivative and
\begin{equation}
\begin{split}
    \Gamma_{\mu\nu}
=& -R\bigl(\nabla_{\mu}\psi_{\nu}+\nabla_{\nu}\psi_{\mu}\bigr)-4\nabla_{\alpha}\psi^{\alpha}\Bigl(R_{\mu\nu}-\tfrac{1}{2}R\,g_{\mu\nu}\Bigr)\\&
   + 4R_{\mu}{}^{\alpha}\nabla_{\alpha}\psi_{\nu} + 4R_{\nu}{}^{\alpha}\nabla_{\alpha}\psi_{\mu} \\&- 4g_{\mu\nu}\,R_{\alpha\beta}\nabla^{\alpha}\psi^{\beta}
   + 4R^{\beta}{}_{\mu\alpha\nu}\,\nabla^{\alpha}\psi_{\beta},
   \end{split}
\end{equation}
with
\begin{equation}
\psi_{\mu} = \lambda^{2}\,f'(\varphi)\,\nabla_{\mu}\varphi.
\end{equation}
Since we are interested in static solutions, we can employ the following ansatz for the metric tensor
\begin{equation}
{\rm d}s^{2} = - e^{2A(r)}\, {\rm d}t^{2} + e^{2B(r)}\, {\rm d}r^{2} 
+ r^{2}{\rm d}\Omega_{S^2}\,,
\end{equation}
where ${\rm d}\Omega_{S^2}$ is the line-element on the Euclidian two-sphere. The relevant components of the equations of motion can then be expressed as
\begin{equation}
\begin{split}
&\frac{1}{r}\!\left[
1 + \frac{2}{r}\bigl(1 - 3e^{-2B}\bigr)\psi_{r}
\right]B'
+ \frac{e^{2B} - 1}{r^{2}}\\
&\qquad\qquad 
- \frac{4}{r^{2}}\bigl(1 - e^{-2B}\bigr)\psi_r'
- \left(\varphi'\right)^{2}
= 0\,,
\\
&\frac{2}{r}\!\left[
1 + \frac{2}{r}\bigl(1 - 3e^{-2B}\bigr)\psi_{r}
\right]A'
- \frac{e^{2B} - 1}{r^{2}}
- \left(\varphi'\right)^{2}
= 0,
\\[1em]
&A''
+ \!\left(A' + \frac{1}{r}\right)
  \!\left(A' - B'\right)\\&\qquad\qquad
+ \frac{4e^{-2B}}{r}\!\left[
  3A'B'
  - A''
  - \!\left(A'\right)^{2}
\right]\!\psi_{r}
\\
&\qquad\qquad
- \frac{4e^{-2B}}{r}
  A'\psi_r'
+ \left(\varphi'\right)^{2}
= 0,
\\
&\varphi''
+ \!\left(
A'-B' + \frac{2}{r}
\right)\!\varphi'\\
&\qquad\qquad
- \frac{2\lambda^{2}}{r^{2}}
  f'(\varphi)
  \Biggl\{
  (1 - e^{-2B})
  \!\left[
  A''
  + A'
    \!\left(A' - B'\right)
  \right]
\\
&\quad\qquad
+ 2e^{-2B}
  A'B'
  \Biggr\}
= 0 \,,
\end{split}
\end{equation}
with $\psi_r=\lambda^2f'(\varphi)\varphi'$. 
 Notice that we employed the symbol $'$ for indicating both the derivative of $f(\varphi)$ with respect to $\varphi$, and the derivative of $A,\,B,\,\varphi,\,\psi_r$ with respect to $r$. This set of equations can be solved imposing asymptotic flatness and regularity of the scalar field at the BH event horizon, which amounts to requiring
 \begin{equation}
\begin{split}
    &A\left.\right|_{r\rightarrow\infty}\rightarrow 0\,,\quad B\left.\right|_{r\rightarrow\infty}\rightarrow 0\,\quad\,,\varphi\left.\right|_{r\rightarrow\infty}\rightarrow 0\,\quad\,, \\
    &e^{2 A}\left.\right|_{r\rightarrow r_{\rm EH}}\rightarrow 0\,,e^{-2B}\left.\right|_{r\rightarrow r_{\rm EH}}\rightarrow 0\,.
\end{split}
 \end{equation}
 Finally, the regularity of the scalar field at the event horizon yields the following boundary condition for its first derivative
 \begin{equation}
     \varphi'(r_{\rm EH})
= 
\frac{r_{\rm EH}}{4\lambda^{2}}\,f'(\varphi_{\rm EH})
\left[
-1 +
\sqrt{
1 - \tfrac{24\lambda^{4}}{r_{{\rm EH}}^{4}}
\left(f'(\varphi_{{\rm EH}})\right)^{2}
}
\right] \,.
 \end{equation}

\section{Comparison with the weak field expansion}
\label{Comparison_Juliè}

\begin{figure}[t]
          \centering
\includegraphics[width =0.48\textwidth]{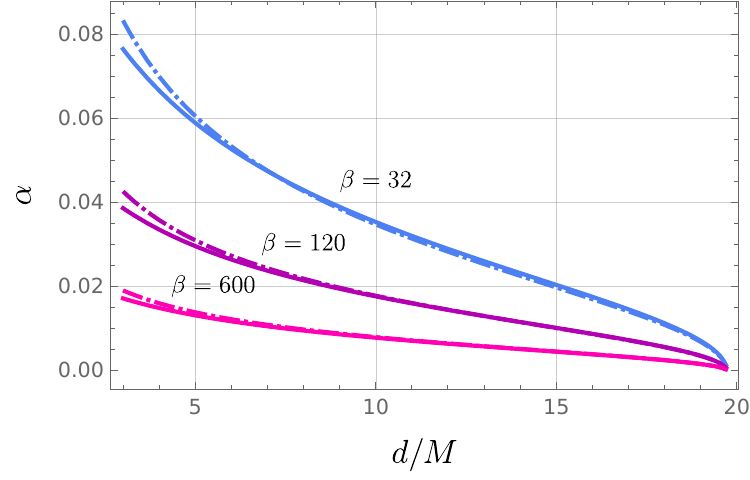}
 \caption{Comparison of the prediction for the dimensionless scalar charge $\alpha$, obtained with the semi-analytic approach described in the main text (continuous lines) and with the perturbative approach of~\cite{Julie:2023ncq} (dot-dashed lines). The curves are represented as a function of the separation for $(\lambda/M)^2=0.703$, and for different values of $\beta$.  
}
\label{fig:scalarization_perturbative}
\end{figure}
In this section, we compare the result on DS obtained by numerically solving Eq.~\eqref{algebraic_system} with the analytic approach derived in~\cite{Julie:2023ncq}.

In the small scalar field limit, the model of Eq.~\eqref{eq:coupling} boils down to 
\begin{equation}
    f(\varphi)=\frac{\varphi^2}{2}-\frac{\beta}{4}\varphi^4
+\mathcal{O}(\varphi^6)\,.
\end{equation}
One can define the following dimensionless scalar charge for an adiabatic sequence of BH solutions, with mass $m$
\begin{equation}
    \alpha \equiv \left.\frac{\partial \ln m}{\partial \varphi}\right|_{\mathcal{S_{\text{W}}}}=-\frac{Q}{m}\,,
\end{equation}
where the second relation follows from the first law of BH mechanics.

Taylor-expanding this quantity about $\varphi = 0$, one gets
\begin{equation}
    \alpha = s_1 \varphi + \frac{1}{6}s_2 \varphi^3+\mathcal{O}(\varphi^5)\,.
\end{equation}
The coefficients $s_i$ of this expansion are called sensitivities. The fact that this expansion only includes odd terms, comes from the $\mathcal{Z}_2$ symmetry of the model.

The sensitivities can be extracted from the numerical sequences of solutions, by fitting them with a cubic polynomial near $\varphi = 0$, and computing the first and third derivatives. 
For an equal-mass binary system, the scalarization radius can be estimated as $d_{\rm DS}\simeq|s_1|/2$. Moreover, the dependence of the dimensionless scalar charge on the separation can be obtained perturbatively in $d-d_{\rm DS}$. At leading order, one has~\cite{Julie:2023ncq} 
\begin{equation}
    \alpha \simeq  -s_1^2\sqrt{\dfrac{3}{3 s_2+s_1^2}}\left(\dfrac{1}{d}-\dfrac{1}{d_{\rm DS}}\right)^{1/2}\,.
    \label{dimensionless_scalar_charge_separation}
\end{equation}
The comparison between this analytic estimate and the prediction obtained through the numerical solution of the algebraic system of Eq.~\eqref{algebraic_system} is shown in  Fig.~\ref{fig:scalarization_perturbative}. The continuous curves represent the non-perturbative solution, while the dot-dashed lines represent the result from Eq.~\eqref{dimensionless_scalar_charge_separation}. The different colors indicate different values of $\beta$, while the dimensional coupling has been fixed at $(\lambda/M)^2 = 0.703$. The comparison shows excellent agreement up to very small separations.

\section{Convergence}
\label{Convergence}

\begin{figure}[b]
    \centering
    \includegraphics[width=0.48\textwidth]{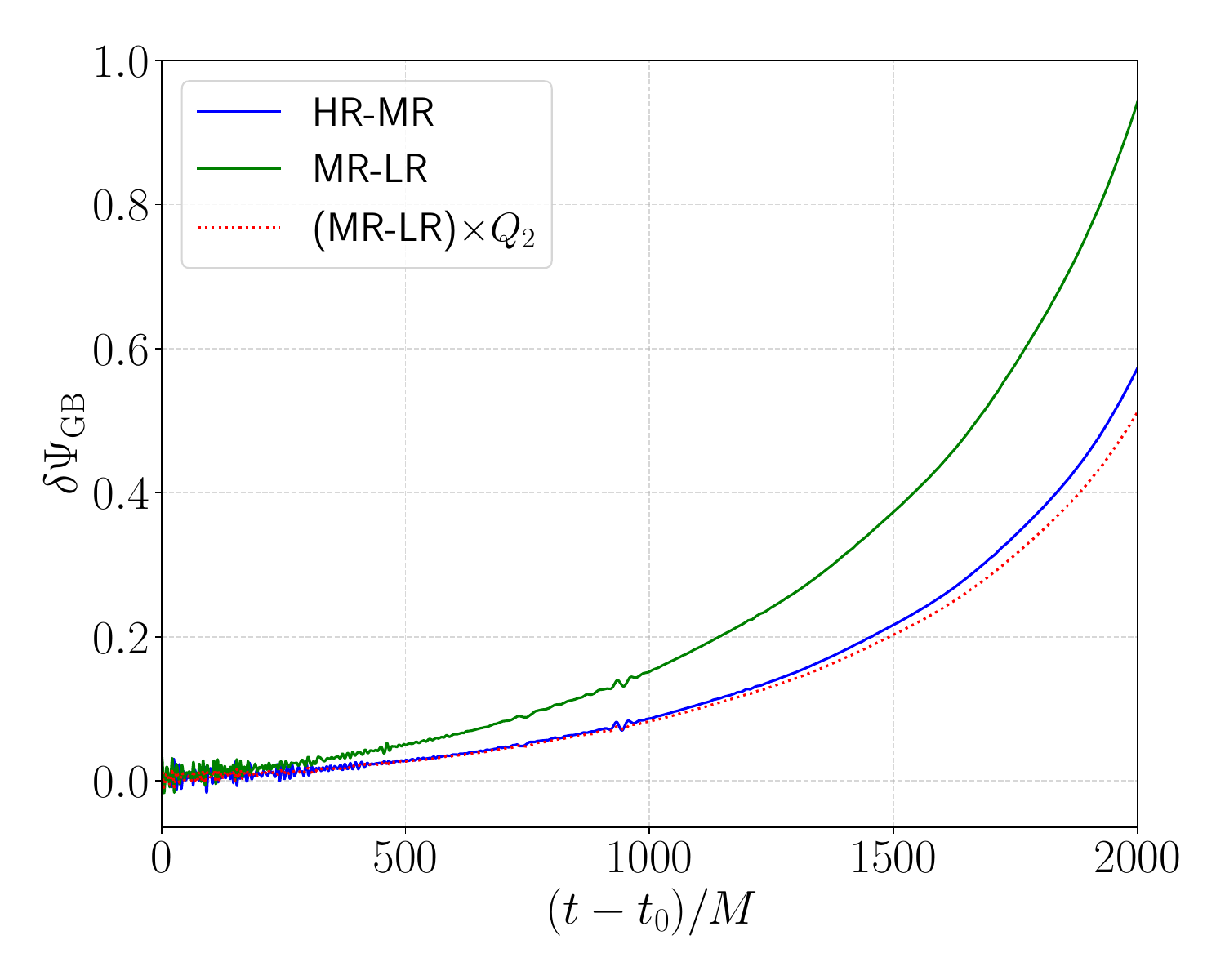}
    \caption{Differences of the orbital phase $\Psi$ for three different resolutions: $N=128$ (low), $N=160$ (medium), and $N=192$ (high). The simulation parameters are $(\lambda/M)^2=0.703$ and $\beta=48$. The red dotted line displays the difference between the low and the medium resolutions, multiplied by a second-order convergence factor $Q_2=\frac{h_{\rm MR}^2-h_{\rm HR}^2}{h_{\rm LR}^2-h_{\rm MR}^2}$, where $h_i$ is the grid spacing for the different resolutions. We choose to do a shift of $t_0=500\,M$, which is needed to avoid the inaccuracies in the phase determination due to the initial gauge settling and abrupt scalar field development.}
    \label{fig:conv}
\end{figure}
We explored the convergence of our simulations by examining $\Psi$, defined as half the complex phase of the $(2,2)$ mode of the Weyl scalar, extracted at $r=140M$. Three different resolutions are considered: $N=128$ (low), $N=160$ (medium), and $N=192$ (high). We have chosen a binary merger simulation with $(\lambda/M)^2=0.703$ and $\beta=48$, starting from a separation of $d/M=13$, corresponding to the $8^{\rm th}$ row in Table \ref{tab:sim_params}. The dephasing between the three resolutions is shown in Fig.~\ref{fig:conv}. The time $t_0$ denotes the beginning of the orbital phase calculation, which is taken to be $500M$ to avoid the influence of initial junk radiation during the gauge settling. As one can see, the results match well a second-order convergence. Note that this corresponds to a lower order of convergence than the one observed in~\cite{AresteSalo:2025sxc}, which is due to the fact that we start with non-constraint satisfying initial data. However, this coincides with the same order of convergence that we would attain if we had solved the constraints using the initial data solver \texttt{GRTresna}~\cite{Aurrekoetxea:2025kmm}. Therefore, we can guarantee that our simulations are stable and in the convergent regime.

\bibliographystyle{ieeetr}
\bibliography{Bibliography}

@article{Doneva2022BeyondTS,
  title={Beyond the spontaneous scalarization: New fully nonlinear mechanism for the formation of scalarized black holes and its dynamical development},
  author={Daniela D. Doneva and Stoytcho S. Yazadjiev},
  journal={Physical Review D},
  year={2022},
  url={https://api.semanticscholar.org/CorpusID:247074698}
}

@article{Danchev:2021tew,
    author = "Danchev, Victor I. and Doneva, Daniela D. and Yazadjiev, Stoytcho S.",
    title = "{Constraining scalarization in scalar-Gauss-Bonnet gravity through binary pulsars}",
    eprint = "2112.03869",
    archivePrefix = "arXiv",
    primaryClass = "gr-qc",
    doi = "10.1103/PhysRevD.106.124001",
    journal = "Phys. Rev. D",
    volume = "106",
    number = "12",
    pages = "124001",
    year = "2022"
}

@article{Wong:2022wni,
    author = "Wong, Leong Khim and Herdeiro, Carlos A. R. and Radu, Eugen",
    title = "{Constraining spontaneous black hole scalarization in scalar-tensor-Gauss-Bonnet theories with current gravitational-wave data}",
    eprint = "2204.09038",
    archivePrefix = "arXiv",
    primaryClass = "gr-qc",
    doi = "10.1103/PhysRevD.106.024008",
    journal = "Phys. Rev. D",
    volume = "106",
    number = "2",
    pages = "024008",
    year = "2022"
}

@article{Taniguchi:2014fqa,
    author = "Taniguchi, Keisuke and Shibata, Masaru and Buonanno, Alessandra",
    title = "{Quasiequilibrium sequences of binary neutron stars undergoing dynamical scalarization}",
    eprint = "1410.0738",
    archivePrefix = "arXiv",
    primaryClass = "gr-qc",
    doi = "10.1103/PhysRevD.91.024033",
    journal = "Phys. Rev. D",
    volume = "91",
    number = "2",
    pages = "024033",
    year = "2015"
}

@article{Chamberlain:2017fjl,
    author = "Chamberlain, Katie and Yunes, Nicolas",
    title = "{Theoretical Physics Implications of Gravitational Wave Observation with Future Detectors}",
    eprint = "1704.08268",
    archivePrefix = "arXiv",
    primaryClass = "gr-qc",
    doi = "10.1103/PhysRevD.96.084039",
    journal = "Phys. Rev. D",
    volume = "96",
    number = "8",
    pages = "084039",
    year = "2017"
}

@article{Herdeiro:2015waa,
    author = "Herdeiro, Carlos A. R. and Radu, Eugen",
    editor = "Herdeiro, Carlos A. R. and Cardoso, Vitor and Lemos, Jose P. S. and Mena, Filipe C.",
    title = "{Asymptotically flat black holes with scalar hair: a review}",
    eprint = "1504.08209",
    archivePrefix = "arXiv",
    primaryClass = "gr-qc",
    doi = "10.1142/S0218271815420146",
    journal = "Int. J. Mod. Phys. D",
    volume = "24",
    number = "09",
    pages = "1542014",
    year = "2015"
}

@article{Yordanov:2024lfk,
    author = "Yordanov, Petar Y. and Staykov, Kalin V. and Yazadjiev, Stoytcho S. and Doneva, Daniela D.",
    title = "{The power of binary pulsars in testing Gauss-Bonnet gravity}",
    eprint = "2402.06305",
    archivePrefix = "arXiv",
    primaryClass = "gr-qc",
    doi = "10.1051/0004-6361/202449679",
    journal = "Astron. Astrophys.",
    volume = "687",
    pages = "A17",
    year = "2024"
}

@article{Lyu:2022gdr,
    author = "Lyu, Zhenwei and Jiang, Nan and Yagi, Kent",
    title = "{Constraints on Einstein-dilation-Gauss-Bonnet gravity from black hole-neutron star gravitational wave events}",
    eprint = "2201.02543",
    archivePrefix = "arXiv",
    primaryClass = "gr-qc",
    reportNumber = "LIGO-P2100466",
    doi = "10.1103/PhysRevD.105.064001",
    journal = "Phys. Rev. D",
    volume = "105",
    number = "6",
    pages = "064001",
    year = "2022",
    note = "[Erratum: Phys.Rev.D 106, 069901 (2022), Erratum: Phys.Rev.D 106, 069901 (2022)]"
}

@article{Yazadjiev:2025ezx,
    author = "Yazadjiev, Stoytcho S. and Doneva, Daniela D.",
    title = "{No-hair theorems in general relativity and scalar{\textendash}tensor theories}",
    eprint = "2505.01038",
    archivePrefix = "arXiv",
    primaryClass = "gr-qc",
    doi = "10.1142/S0218271825300046",
    journal = "Int. J. Mod. Phys. D",
    volume = "34",
    number = "09",
    pages = "2530004",
    year = "2025"
}

@article{Doneva:2017bvd,
    author = "Doneva, Daniela D. and Yazadjiev, Stoytcho S.",
    title = "{New Gauss-Bonnet Black Holes with Curvature-Induced Scalarization in Extended Scalar-Tensor Theories}",
    eprint = "1711.01187",
    archivePrefix = "arXiv",
    primaryClass = "gr-qc",
    doi = "10.1103/PhysRevLett.120.131103",
    journal = "Phys. Rev. Lett.",
    volume = "120",
    number = "13",
    pages = "131103",
    year = "2018"
}

@article{Blazquez-Salcedo:2018jnn,
    author = "Bl{\'a}zquez-Salcedo, Jose Luis and Doneva, Daniela D. and Kunz, Jutta and Yazadjiev, Stoytcho S.",
    title = "{Radial perturbations of the scalarized Einstein-Gauss-Bonnet black holes}",
    eprint = "1805.05755",
    archivePrefix = "arXiv",
    primaryClass = "gr-qc",
    doi = "10.1103/PhysRevD.98.084011",
    journal = "Phys. Rev. D",
    volume = "98",
    number = "8",
    pages = "084011",
    year = "2018"
}

@article{Silva:2017uqg,
    author = "Silva, Hector O. and Sakstein, Jeremy and Gualtieri, Leonardo and Sotiriou, Thomas P. and Berti, Emanuele",
    title = "{Spontaneous scalarization of black holes and compact stars from a Gauss-Bonnet coupling}",
    eprint = "1711.02080",
    archivePrefix = "arXiv",
    primaryClass = "gr-qc",
    doi = "10.1103/PhysRevLett.120.131104",
    journal = "Phys. Rev. Lett.",
    volume = "120",
    number = "13",
    pages = "131104",
    year = "2018"
}

@article{Julie:2023ncq,
    author = "Juli{\'e}, F{\'e}lix-Louis",
    title = "{Dynamical scalarization in Schwarzschild binary inspirals}",
    eprint = "2312.16764",
    archivePrefix = "arXiv",
    primaryClass = "gr-qc",
    month = "12",
    year = "2023"
}

@article{Julie:2022huo,
    author = "Juli\'e, F\'elix-Louis and Silva, Hector O. and Berti, Emanuele and Yunes, Nicol\'as",
    title = "{Black hole sensitivities in Einstein-scalar-Gauss-Bonnet gravity}",
    journal = "Phys. Rev. D",
    year = "2022"
}

@article{Barausse_2013,
   title={Neutron-star mergers in scalar-tensor theories of gravity},
   volume={87},
   ISSN={1550-2368},
   url={http://dx.doi.org/10.1103/PhysRevD.87.081506},
   DOI={10.1103/physrevd.87.081506},
   number={8},
   journal={Physical Review D},
   publisher={American Physical Society (APS)},
   author={Barausse, Enrico and Palenzuela, Carlos and Ponce, Marcelo and Lehner, Luis},
   year={2013},
   month=apr }

@article{Palenzuela_2014,
   title={Dynamical scalarization of neutron stars in scalar-tensor gravity theories},
   url={http://dx.doi.org/10.1103/PhysRevD.89.044024},
   journal={Physical Review D},
   publisher={American Physical Society (APS)},
   author={Palenzuela, Carlos and Barausse, Enrico and Ponce, Marcelo and Lehner, Luis},
   year={2014},
  }

@article{Sampson_2014,
   title={Projected constraints on scalarization with gravitational waves from neutron star binaries},
   volume={90},
   url={http://dx.doi.org/10.1103/PhysRevD.90.124091},
   journal={Physical Review D},
   publisher={American Physical Society (APS)},
   author={Sampson, Laura and Yunes, Nicolás and Cornish, Neil and Ponce, Marcelo and Barausse, Enrico and Klein, Antoine and Palenzuela, Carlos and Lehner, Luis},
   year={2014},
}

@article{Shibata:2013pra,
    author = "Shibata, Masaru and Taniguchi, Keisuke and Okawa, Hirotada and Buonanno, Alessandra",
    title = "{Coalescence of binary neutron stars in a scalar-tensor theory of gravity}",
    eprint = "1310.0627",
    archivePrefix = "arXiv",
    primaryClass = "gr-qc",
    doi = "10.1103/PhysRevD.89.084005",
    journal = "Phys. Rev. D",
    volume = "89",
    number = "8",
    pages = "084005",
    year = "2014"
}

@article{Kuan:2023trn,
    author = "Kuan, Hao-Jui and Lam, Alan Tsz-Lok and Doneva, Daniela D. and Yazadjiev, Stoytcho S. and Shibata, Masaru and Kiuchi, Kenta",
    title = "{Dynamical scalarization during neutron star mergers in scalar-Gauss-Bonnet theory}",
    eprint = "2302.11596",
    archivePrefix = "arXiv",
    primaryClass = "gr-qc",
    doi = "10.1103/PhysRevD.108.063033",
    journal = "Phys. Rev. D",
    volume = "108",
    number = "6",
    pages = "063033",
    year = "2023"
}

@misc{doneva2022scalarization,
      title={Scalarization}, 
      author={Daniela D. Doneva and Fethi M. Ramazanoğlu and Hector O. Silva and Thomas P. Sotiriou and Stoytcho S. Yazadjiev},
      year={2022},
}

@article{Capuano:2023yyh,
    author = "Capuano, Lodovico and Santoni, Luca and Barausse, Enrico",
    title = "{Black hole hairs in scalar-tensor gravity and the lack thereof}",
    eprint = "2304.12750",
    archivePrefix = "arXiv",
    primaryClass = "gr-qc",
    reportNumber = "DE13253",
    doi = "10.1103/PhysRevD.108.064058",
    journal = "Phys. Rev. D",
    volume = "108",
    number = "6",
    pages = "064058",
    year = "2023"
}

@article{Hui:2012qt,
    author = "Hui, Lam and Nicolis, Alberto",
    title = "{No-Hair Theorem for the Galileon}",
    eprint = "1202.1296",
    archivePrefix = "arXiv",
    primaryClass = "hep-th",
    doi = "10.1103/PhysRevLett.110.241104",
    journal = "Phys. Rev. Lett.",
    volume = "110",
    pages = "241104",
    year = "2013"
}

@article{Graham:2014mda,
    author = "Graham, Alexander A. H. and Jha, Rahul",
    title = "{Nonexistence of black holes with noncanonical scalar fields}",
    eprint = "1401.8203",
    archivePrefix = "arXiv",
    primaryClass = "gr-qc",
    doi = "10.1103/PhysRevD.89.084056",
    journal = "Phys. Rev. D",
    volume = "89",
    number = "8",
    pages = "084056",
    year = "2014",
    note = "[Erratum: Phys.Rev.D 92, 069901 (2015)]"
}

@article{Bekenstein:1971hc,
    author = "Bekenstein, Jacob D.",
    title = "{Nonexistence of baryon number for static black holes}",
    doi = "10.1103/PhysRevD.5.1239",
    journal = "Phys. Rev. D",
    volume = "5",
    pages = "1239--1246",
    year = "1972"
}

@article{Bekenstein:1972ky,
    author = "Bekenstein, J. D.",
    title = "{Nonexistence of baryon number for black holes. ii}",
    doi = "10.1103/PhysRevD.5.2403",
    journal = "Phys. Rev. D",
    volume = "5",
    pages = "2403--2412",
    year = "1972"
}

@article{Fierz:1939ix,
    author = "Fierz, M. and Pauli, W.",
    title = "{On relativistic wave equations for particles of arbitrary spin in an electromagnetic field}",
    doi = "10.1098/rspa.1939.0140",
    journal = "Proc. Roy. Soc. Lond. A",
    volume = "173",
    pages = "211--232",
    year = "1939"
}

@article{Brans:1961sx,
    author = "Brans, C. and Dicke, R. H.",
    editor = "Hsu, Jong-Ping and Fine, D.",
    title = "{Mach's principle and a relativistic theory of gravitation}",
    doi = "10.1103/PhysRev.124.925",
    journal = "Phys. Rev.",
    volume = "124",
    pages = "925--935",
    year = "1961"
}

@article{Jordan:1959eg,
    author = "Jordan, Pascual",
    title = "{The present state of Dirac's cosmological hypothesis}",
    doi = "10.1007/BF01375155",
    journal = "Z. Phys.",
    volume = "157",
    pages = "112--121",
    year = "1959"
}

@article{Dicke:1961gz,
    author = "Dicke, R. H.",
    title = "{Mach's principle and invariance under transformation of units}",
    doi = "10.1103/PhysRev.125.2163",
    journal = "Phys. Rev.",
    volume = "125",
    pages = "2163--2167",
    year = "1962"
}

@article{Damour:1992we,
    author = "Damour, Thibault and Esposito-Farese, Gilles",
    title = "{Tensor multiscalar theories of gravitation}",
    reportNumber = "IHES-P-91-93, CPT-91-PE-2542",
    doi = "10.1088/0264-9381/9/9/015",
    journal = "Class. Quant. Grav.",
    volume = "9",
    pages = "2093--2176",
    year = "1992"
}

@article{Damour:1993hw,
    author = "Damour, Thibault and Esposito-Farese, Gilles",
    title = "{Nonperturbative strong field effects in tensor - scalar theories of gravitation}",
    reportNumber = "IHES-P-93-1, CPT-93-PE-2868",
    doi = "10.1103/PhysRevLett.70.2220",
    journal = "Phys. Rev. Lett.",
    volume = "70",
    pages = "2220--2223",
    year = "1993"
}

@article{Horndeski:1974wa,
    author = "Horndeski, Gregory Walter",
    title = "{Second-order scalar-tensor field equations in a four-dimensional space}",
    doi = "10.1007/BF01807638",
    journal = "Int. J. Theor. Phys.",
    volume = "10",
    pages = "363--384",
    year = "1974"
}

@article{Gleyzes:2014dya,
    author = "Gleyzes, J\'er\^ome and Langlois, David and Piazza, Federico and Vernizzi, Filippo",
    title = "{Healthy theories beyond Horndeski}",
    eprint = "1404.6495",
    archivePrefix = "arXiv",
    primaryClass = "hep-th",
    doi = "10.1103/PhysRevLett.114.211101",
    journal = "Phys. Rev. Lett.",
    volume = "114",
    number = "21",
    pages = "211101",
    year = "2015"
}

@inproceedings{Langlois:2017mdk,
    author = "Langlois, David",
    title = "{Degenerate Higher-Order Scalar-Tensor (DHOST) theories}",
    booktitle = "{52nd Rencontres de Moriond on Gravitation}",
    eprint = "1707.03625",
    archivePrefix = "arXiv",
    primaryClass = "gr-qc",
    pages = "221--228",
    year = "2017"
}

@article{Wald:1993nt,
    author = "Wald, Robert M.",
    title = "{Black hole entropy is the Noether charge}",
    eprint = "gr-qc/9307038",
    archivePrefix = "arXiv",
    reportNumber = "EFI-93-42",
    doi = "10.1103/PhysRevD.48.R3427",
    journal = "Phys. Rev. D",
    volume = "48",
    number = "8",
    pages = "R3427--R3431",
    year = "1993"
}

@article{Iyer:1994ys,
    author = "Iyer, Vivek and Wald, Robert M.",
    title = "{Some properties of Noether charge and a proposal for dynamical black hole entropy}",
    eprint = "gr-qc/9403028",
    archivePrefix = "arXiv",
    doi = "10.1103/PhysRevD.50.846",
    journal = "Phys. Rev. D",
    volume = "50",
    pages = "846--864",
    year = "1994"
}

@article{Shiralilou:2021mfl,
    author = "Shiralilou, Banafsheh and Hinderer, Tanja and Nissanke, Samaya M. and Ortiz, N\'estor and Witek, Helvi",
    title = "{Post-Newtonian gravitational and scalar waves in scalar-Gauss\textendash{}Bonnet gravity}",
    eprint = "2105.13972",
    archivePrefix = "arXiv",
    primaryClass = "gr-qc",
    doi = "10.1088/1361-6382/ac4196",
    journal = "Class. Quant. Grav.",
    volume = "39",
    number = "3",
    pages = "035002",
    year = "2022"
}

@article{LIGOScientific:2021sio,
    author = "Abbott, R. and others",
    collaboration = "LIGO Scientific, VIRGO, KAGRA",
    title = "{Tests of General Relativity with GWTC-3}",
    eprint = "2112.06861",
    archivePrefix = "arXiv",
    primaryClass = "gr-qc",
    reportNumber = "LIGO-P2100275",
    month = "12",
    year = "2021"
}

@article{KAGRA:2021duu,
    author = "Abbott, R. and others",
    collaboration = "KAGRA, VIRGO, LIGO Scientific",
    title = "{Population of Merging Compact Binaries Inferred Using Gravitational Waves through GWTC-3}",
    eprint = "2111.03634",
    archivePrefix = "arXiv",
    primaryClass = "astro-ph.HE",
    reportNumber = "LIGO-P2100239 ; Data release: https://zenodo.org/record/5655785, LIGO-P2100239",
    doi = "10.1103/PhysRevX.13.011048",
    journal = "Phys. Rev. X",
    volume = "13",
    number = "1",
    pages = "011048@article{Akyuz:2025seg,
    author = {Aky{\"u}z, Aleyna and Correia, Alex and Garofalo, Jada and Kacanja, Keisi and Roy, Labani and Soni, Kanchan and Tan, Hung and Y, Vikas Jadhav and Nitz, Alexander H. and Capano, Collin D.},
    title = "{Potential science with GW250114 -- the loudest binary black hole merger detected to date}",
    eprint = "2507.08789",
    archivePrefix = "arXiv",
    primaryClass = "gr-qc",
    month = "7",
    year = "2025"
}",
    year = "2023"
}

@article{Akyuz:2025seg,
    author = {Aky{\"u}z, Aleyna and Correia, Alex and Garofalo, Jada and Kacanja, Keisi and Roy, Labani and Soni, Kanchan and Tan, Hung and Y, Vikas Jadhav and Nitz, Alexander H. and Capano, Collin D.},
    title = "{Potential science with GW250114 -- the loudest binary black hole merger detected to date}",
    eprint = "2507.08789",
    archivePrefix = "arXiv",
    primaryClass = "gr-qc",
    month = "7",
    year = "2025"
}

@article{KAGRA:2025oiz,
    author = "Abac, A. G. and others",
    collaboration = "KAGRA, Virgo, LIGO Scientific",
    title = "{GW250114: Testing Hawking{\textquoteright}s Area Law and the Kerr Nature of Black Holes}",
    eprint = "2509.08054",
    archivePrefix = "arXiv",
    primaryClass = "gr-qc",
    reportNumber = "LIGO-P2500421",
    doi = "10.1103/kw5g-d732",
    journal = "Phys. Rev. Lett.",
    volume = "135",
    number = "11",
    pages = "111403",
    year = "2025"
}

@article{LIGOScientific:2025obp,
    author = "Abac, A. G. and others",
    collaboration = "LIGO Scientific, VIRGO, KAGRA",
    title = "{Black Hole Spectroscopy and Tests of General Relativity with GW250114}",
    eprint = "2509.08099",
    archivePrefix = "arXiv",
    primaryClass = "gr-qc",
    reportNumber = "LIGO P2500461",
    month = "9",
    year = "2025"
}

@article{KAGRA:2021vkt,
    author = "Abbott, R. and others",
    collaboration = "KAGRA, VIRGO, LIGO Scientific",
    title = "{GWTC-3: Compact Binary Coalescences Observed by LIGO and Virgo during the Second Part of the Third Observing Run}",
    eprint = "2111.03606",
    archivePrefix = "arXiv",
    primaryClass = "gr-qc",
    reportNumber = "LIGO-P2000318",
    doi = "10.1103/PhysRevX.13.041039",
    journal = "Phys. Rev. X",
    volume = "13",
    number = "4",
    pages = "041039",
    year = "2023"
}

@article{LIGOScientific:2020ibl,
    author = "Abbott, R. and others",
    collaboration = "LIGO Scientific, Virgo",
    title = "{GWTC-2: Compact Binary Coalescences Observed by LIGO and Virgo During the First Half of the Third Observing Run}",
    eprint = "2010.14527",
    archivePrefix = "arXiv",
    primaryClass = "gr-qc",
    reportNumber = "P2000061",
    doi = "10.1103/PhysRevX.11.021053",
    journal = "Phys. Rev. X",
    volume = "11",
    pages = "021053",
    year = "2021"
}

@article{LIGOScientific:2021usb,
    author = "Abbott, R. and others",
    collaboration = "LIGO Scientific, VIRGO",
    title = "{GWTC-2.1: Deep extended catalog of compact binary coalescences observed by LIGO and Virgo during the first half of the third observing run}",
    eprint = "2108.01045",
    archivePrefix = "arXiv",
    primaryClass = "gr-qc",
    reportNumber = "LIGO-P2100063",
    doi = "10.1103/PhysRevD.109.022001",
    journal = "Phys. Rev. D",
    volume = "109",
    number = "2",
    pages = "022001",
    year = "2024"
}

@article{LIGOScientific:2021qlt,
    author = "Abbott, R. and others",
    collaboration = "LIGO Scientific, KAGRA, VIRGO",
    title = "{Observation of Gravitational Waves from Two Neutron Star{\textendash}Black Hole Coalescences}",
    eprint = "2106.15163",
    archivePrefix = "arXiv",
    primaryClass = "astro-ph.HE",
    reportNumber = "LIGO Document P2000357",
    doi = "10.3847/2041-8213/ac082e",
    journal = "Astrophys. J. Lett.",
    volume = "915",
    number = "1",
    pages = "L5",
    year = "2021"
}

@article{LIGOScientific:2025slb,
    author = "Abac, A. G. and others",
    collaboration = "LIGO Scientific, VIRGO, KAGRA",
    title = "{GWTC-4.0: Updating the Gravitational-Wave Transient Catalog with Observations from the First Part of the Fourth LIGO-Virgo-KAGRA Observing Run}",
    eprint = "2508.18082",
    archivePrefix = "arXiv",
    primaryClass = "gr-qc",
    reportNumber = "LIGO-P2400386",
    month = "8",
    year = "2025"
}

@article{Wang:2024erh,
    author = "Wang, Xian-Liang and Yang, Shu-Cheng and Han, Wen-Biao",
    title = "{Tests of gravitational wave propagation with LIGO-Virgo catalog}",
    eprint = "2404.14684",
    archivePrefix = "arXiv",
    primaryClass = "gr-qc",
    month = "4",
    year = "2024"
}

@article{AresteSalo:2023hcp,
    author = "Arest\'e Sal\'o, Llibert and Brady, Sam E. and Clough, Katy and Doneva, Daniela and Evstafyeva, Tamara and Figueras, Pau and Fran\c{c}a, Tiago and Rossi, Lorenzo and Yao, Shunhui",
    title = "{GRFolres: A code for modified gravity simulations in strong gravity}",
    eprint = "2309.06225",
    archivePrefix = "arXiv",
    primaryClass = "gr-qc",
    doi = "10.21105/joss.06369",
    journal = "J. Open Source Softw.",
    volume = "9",
    number = "98",
    pages = "6369",
    year = "2024"
}

@article{Andrade:2021rbd,
    author = "Andrade, Tomas and others",
    title = "{GRChombo: An adaptable numerical relativity code for fundamental physics}",
    eprint = "2201.03458",
    archivePrefix = "arXiv",
    primaryClass = "gr-qc",
    doi = "10.21105/joss.03703",
    journal = "J. Open Source Softw.",
    volume = "6",
    number = "68",
    pages = "3703",
    year = "2021"
}

@article{Radia:2021smk,
    author = "Radia, Miren and Sperhake, Ulrich and Drew, Amelia and Clough, Katy and Figueras, Pau and Lim, Eugene A. and Ripley, Justin L. and Aurrekoetxea, Josu C. and Fran\c{c}a, Tiago and Helfer, Thomas",
    title = "{Lessons for adaptive mesh refinement in numerical relativity}",
    eprint = "2112.10567",
    archivePrefix = "arXiv",
    primaryClass = "gr-qc",
    reportNumber = "KCL-PH-TH/2021-89",
    doi = "10.1088/1361-6382/ac6fa9",
    journal = "Class. Quant. Grav.",
    volume = "39",
    number = "13",
    pages = "135006",
    year = "2022"
}

@article{Ansorg:2004ds,
    author = "Ansorg, Marcus and Br{\"u}gmann, Bernd and Tichy, Wolfgang",
    title = "{A Single-domain spectral method for black hole puncture data}",
    eprint = "gr-qc/0404056",
    archivePrefix = "arXiv",
    reportNumber = "CGPG-03-12-3",
    doi = "10.1103/PhysRevD.70.064011",
    journal = "Phys. Rev. D",
    volume = "70",
    pages = "064011",
    year = "2004"
}

@article{Bruegmann:2006ulg,
    author = "Br{\"u}gmann, Bernd and Gonz\'alez, Jos\'e A. and Hannam, Mark and Husa, Sascha and Sperhake, Ulrich and Tichy, Wolfgang",
    title = "{Calibration of Moving Puncture Simulations}",
    eprint = "gr-qc/0610128",
    archivePrefix = "arXiv",
    doi = "10.1103/PhysRevD.77.024027",
    journal = "Phys. Rev. D",
    volume = "77",
    pages = "024027",
    year = "2008"
}

@article{Corman:2022xqg,
    author = "Corman, Maxence and Ripley, Justin L. and East, William E.",
    title = "{Nonlinear studies of binary black hole mergers in Einstein-scalar-Gauss-Bonnet gravity}",
    eprint = "2210.09235",
    archivePrefix = "arXiv",
    primaryClass = "gr-qc",
    doi = "10.1103/PhysRevD.107.024014",
    journal = "Phys. Rev. D",
    volume = "107",
    number = "2",
    pages = "024014",
    year = "2023"
}

@article{AresteSalo:2022hua,
    author = "Arest\'e Sal\'o, Llibert and Clough, Katy and Figueras, Pau",
    title = "{Well-Posedness of the Four-Derivative Scalar-Tensor Theory of Gravity in Singularity Avoiding Coordinates}",
    eprint = "2208.14470",
    archivePrefix = "arXiv",
    primaryClass = "gr-qc",
    doi = "10.1103/PhysRevLett.129.261104",
    journal = "Phys. Rev. Lett.",
    volume = "129",
    number = "26",
    pages = "261104",
    year = "2022"
}

@article{Khalil:2019wyy,
    author = "Khalil, Mohammed and Sennett, Noah and Steinhoff, Jan and Buonanno, Alessandra",
    title = "{Theory-agnostic framework for dynamical scalarization of compact binaries}",
    eprint = "1906.08161",
    archivePrefix = "arXiv",
    primaryClass = "gr-qc",
    doi = "10.1103/PhysRevD.100.124013",
    journal = "Phys. Rev. D",
    volume = "100",
    number = "12",
    pages = "124013",
    year = "2019"
}

@article{Sennett:2017lcx,
    author = "Sennett, Noah and Shao, Lijing and Steinhoff, Jan",
    title = "{Effective action model of dynamically scalarizing binary neutron stars}",
    eprint = "1708.08285",
    archivePrefix = "arXiv",
    primaryClass = "gr-qc",
    doi = "10.1103/PhysRevD.96.084019",
    journal = "Phys. Rev. D",
    volume = "96",
    number = "8",
    pages = "084019",
    year = "2017"
}

@article{Sennett:2016rwa,
    author = "Sennett, Noah and Buonanno, Alessandra",
    title = "{Modeling dynamical scalarization with a resummed post-Newtonian expansion}",
    eprint = "1603.03300",
    archivePrefix = "arXiv",
    primaryClass = "gr-qc",
    doi = "10.1103/PhysRevD.93.124004",
    journal = "Phys. Rev. D",
    volume = "93",
    number = "12",
    pages = "124004",
    year = "2016"
}

@article{AresteSalo:2025sxc,
    author = "Arest{\'e} Sal{\'o}, Llibert and Doneva, Daniela D. and Clough, Katy and Figueras, Pau and Yazadjiev, Stoytcho S.",
    title = "{Challenges in the nonlinear evolution of unequal mass binaries in scalar-Gauss-Bonnet gravity}",
    eprint = "2507.13046",
    archivePrefix = "arXiv",
    primaryClass = "gr-qc",
    doi = "10.1103/tr7v-jhhm",
    journal = "Phys. Rev. D",
    volume = "112",
    number = "8",
    pages = "084022",
    year = "2025"
}

@article{AresteSalo:2023mmd,
    author = "Arest{\'e} Sal{\'o}, Llibert and Clough, Katy and Figueras, Pau",
    title = "{Puncture gauge formulation for Einstein-Gauss-Bonnet gravity and four-derivative scalar-tensor theories in d+1 spacetime dimensions}",
    eprint = "2306.14966",
    archivePrefix = "arXiv",
    primaryClass = "gr-qc",
    doi = "10.1103/PhysRevD.108.084018",
    journal = "Phys. Rev. D",
    volume = "108",
    number = "8",
    pages = "084018",
    year = "2023"
}

@article{Brady:2023dgu,
    author = "Brady, Sam E. and Arest{\'e} Sal{\'o}, Llibert and Clough, Katy and Figueras, Pau and S., Annamalai P.",
    title = "{Solving the initial conditions problem for modified gravity theories}",
    eprint = "2308.16791",
    archivePrefix = "arXiv",
    primaryClass = "gr-qc",
    doi = "10.1103/PhysRevD.108.104022",
    journal = "Phys. Rev. D",
    volume = "108",
    number = "10",
    pages = "104022",
    year = "2023"
}

@article{Nee:2024bur,
    author = "Nee, Peter James and Lara, Guillermo and Pfeiffer, Harald P. and Vu, Nils L.",
    title = "{Quasistationary hair for binary black hole initial data in scalar Gauss-Bonnet gravity}",
    eprint = "2406.08410",
    archivePrefix = "arXiv",
    primaryClass = "gr-qc",
    doi = "10.1103/PhysRevD.111.024061",
    journal = "Phys. Rev. D",
    volume = "111",
    number = "2",
    pages = "024061",
    year = "2025"
}

@article{Lara:2025kzj,
    author = "Lara, Guillermo and others",
    title = "{Signatures from metastable oppositely-charged black hole binaries in scalar Gauss-Bonnet gravity}",
    eprint = "2505.14785",
    archivePrefix = "arXiv",
    primaryClass = "gr-qc",
    month = "5",
    year = "2025"
}

@article{Kovacs:2020pns,
    author = "Kov{\'a}cs, {\'A}ron D. and Reall, Harvey S.",
    title = "{Well-Posed Formulation of Scalar-Tensor Effective Field Theory}",
    eprint = "2003.04327",
    archivePrefix = "arXiv",
    primaryClass = "gr-qc",
    doi = "10.1103/PhysRevLett.124.221101",
    journal = "Phys. Rev. Lett.",
    volume = "124",
    number = "22",
    pages = "221101",
    year = "2020"
}

@article{Kovacs:2020ywu,
    author = "Kov{\'a}cs, {\'A}ron D. and Reall, Harvey S.",
    title = "{Well-posed formulation of Lovelock and Horndeski theories}",
    eprint = "2003.08398",
    archivePrefix = "arXiv",
    primaryClass = "gr-qc",
    doi = "10.1103/PhysRevD.101.124003",
    journal = "Phys. Rev. D",
    volume = "101",
    number = "12",
    pages = "124003",
    year = "2020"
}

@article{East:2020hgw,
    author = "East, William E. and Ripley, Justin L.",
    title = "{Evolution of Einstein-scalar-Gauss-Bonnet gravity using a modified harmonic formulation}",
    eprint = "2011.03547",
    archivePrefix = "arXiv",
    primaryClass = "gr-qc",
    doi = "10.1103/PhysRevD.103.044040",
    journal = "Phys. Rev. D",
    volume = "103",
    number = "4",
    pages = "044040",
    year = "2021"
}

@article{East:2021bqk,
    author = "East, William E. and Ripley, Justin L.",
    title = "{Dynamics of Spontaneous Black Hole Scalarization and Mergers in Einstein-Scalar-Gauss-Bonnet Gravity}",
    eprint = "2105.08571",
    archivePrefix = "arXiv",
    primaryClass = "gr-qc",
    doi = "10.1103/PhysRevLett.127.101102",
    journal = "Phys. Rev. Lett.",
    volume = "127",
    number = "10",
    pages = "101102",
    year = "2021"
}

@article{East:2022rqi,
    author = "East, William E. and Pretorius, Frans",
    title = "{Binary neutron star mergers in Einstein-scalar-Gauss-Bonnet gravity}",
    eprint = "2208.09488",
    archivePrefix = "arXiv",
    primaryClass = "gr-qc",
    doi = "10.1103/PhysRevD.106.104055",
    journal = "Phys. Rev. D",
    volume = "106",
    number = "10",
    pages = "104055",
    year = "2022"
}

@article{Corman:2024vlk,
    author = "Corman, Maxence and East, William E.",
    title = "{Black hole-neutron star mergers in Einstein-scalar-Gauss-Bonnet gravity}",
    eprint = "2405.18496",
    archivePrefix = "arXiv",
    primaryClass = "gr-qc",
    doi = "10.1103/PhysRevD.110.084065",
    journal = "Phys. Rev. D",
    volume = "110",
    number = "8",
    pages = "084065",
    year = "2024"
}

@article{Okounkova:2017yby,
    author = "Okounkova, Maria and Stein, Leo C. and Scheel, Mark A. and Hemberger, Daniel A.",
    title = "{Numerical binary black hole mergers in dynamical Chern-Simons gravity: Scalar field}",
    eprint = "1705.07924",
    archivePrefix = "arXiv",
    primaryClass = "gr-qc",
    doi = "10.1103/PhysRevD.96.044020",
    journal = "Phys. Rev. D",
    volume = "96",
    number = "4",
    pages = "044020",
    year = "2017"
}

@article{Silva:2020omi,
    author = "Silva, Hector O. and Witek, Helvi and Elley, Matthew and Yunes, Nicol{\'a}s",
    title = "{Dynamical Descalarization in Binary Black Hole Mergers}",
    eprint = "2012.10436",
    archivePrefix = "arXiv",
    primaryClass = "gr-qc",
    doi = "10.1103/PhysRevLett.127.031101",
    journal = "Phys. Rev. Lett.",
    volume = "127",
    number = "3",
    pages = "031101",
    year = "2021"
}

@article{Elley:2022ept,
    author = "Elley, Matthew and Silva, Hector O. and Witek, Helvi and Yunes, Nicol{\'a}s",
    title = "{Spin-induced dynamical scalarization, descalarization, and stealthness in scalar-Gauss-Bonnet gravity during a black hole coalescence}",
    eprint = "2205.06240",
    archivePrefix = "arXiv",
    primaryClass = "gr-qc",
    doi = "10.1103/PhysRevD.106.044018",
    journal = "Phys. Rev. D",
    volume = "106",
    number = "4",
    pages = "044018",
    year = "2022"
}

@article{Doneva:2022byd,
    author = "Doneva, Daniela D. and Va{\~n}{\'o}-Vi{\~n}uales, Alex and Yazadjiev, Stoytcho S.",
    title = "{Dynamical descalarization with a jump during a black hole merger}",
    eprint = "2204.05333",
    archivePrefix = "arXiv",
    primaryClass = "gr-qc",
    doi = "10.1103/PhysRevD.106.L061502",
    journal = "Phys. Rev. D",
    volume = "106",
    number = "6",
    pages = "L061502",
    year = "2022"
}

@article{Evstafyeva:2022rve,
    author = "Evstafyeva, Tamara and Agathos, Michalis and Ripley, Justin L.",
    title = "{Measuring the ringdown scalar polarization of gravitational waves in Einstein-scalar-Gauss-Bonnet gravity}",
    eprint = "2212.11359",
    archivePrefix = "arXiv",
    primaryClass = "gr-qc",
    doi = "10.1103/PhysRevD.107.124010",
    journal = "Phys. Rev. D",
    volume = "107",
    number = "12",
    pages = "124010",
    year = "2023"
}

@article{Shiralilou:2020gah,
    author = "Shiralilou, Banafsheh and Hinderer, Tanja and Nissanke, Samaya and Ortiz, N{\'e}stor and Witek, Helvi",
    title = "{Nonlinear curvature effects in gravitational waves from inspiralling black hole binaries}",
    eprint = "2012.09162",
    archivePrefix = "arXiv",
    primaryClass = "gr-qc",
    doi = "10.1103/PhysRevD.103.L121503",
    journal = "Phys. Rev. D",
    volume = "103",
    number = "12",
    pages = "L121503",
    year = "2021"
}

@article{Julie:2019sab,
    author = "Juli{\'e}, F{\'e}lix-Louis and Berti, Emanuele",
    title = "{Post-Newtonian dynamics and black hole thermodynamics in Einstein-scalar-Gauss-Bonnet gravity}",
    eprint = "1909.05258",
    archivePrefix = "arXiv",
    primaryClass = "gr-qc",
    doi = "10.1103/PhysRevD.100.104061",
    journal = "Phys. Rev. D",
    volume = "100",
    number = "10",
    pages = "104061",
    year = "2019"
}

@article{Yagi:2011xp,
    author = "Yagi, Kent and Stein, Leo C. and Yunes, Nicol{\'a}s and Tanaka, Takahiro",
    title = "{Post-Newtonian, Quasi-Circular Binary Inspirals in Quadratic Modified Gravity}",
    eprint = "1110.5950",
    archivePrefix = "arXiv",
    primaryClass = "gr-qc",
    doi = "10.1103/PhysRevD.85.064022",
    journal = "Phys. Rev. D",
    volume = "85",
    pages = "064022",
    year = "2012",
    note = "[Erratum: Phys.Rev.D 93, 029902 (2016)]"
}

@article{Aurrekoetxea:2025kmm,
    author = "Aurrekoetxea, Josu C. and others",
    title = "{GRTresna: An open-source code to solve the initial data constraints in numerical relativity}",
    eprint = "2501.13046",
    archivePrefix = "arXiv",
    primaryClass = "gr-qc",
    month = "1",
    year = "2025"
}

@article{Sotiriou:2013qea,
    author = "Sotiriou, Thomas P. and Zhou, Shuang-Yong",
    title = "{Black hole hair in generalized scalar-tensor gravity}",
    eprint = "1312.3622",
    archivePrefix = "arXiv",
    primaryClass = "gr-qc",
    doi = "10.1103/PhysRevLett.112.251102",
    journal = "Phys. Rev. Lett.",
    volume = "112",
    pages = "251102",
    year = "2014"
}

@article{Sotiriou:2014pfa,
    author = "Sotiriou, Thomas P. and Zhou, Shuang-Yong",
    title = "{Black hole hair in generalized scalar-tensor gravity: An explicit example}",
    eprint = "1408.1698",
    archivePrefix = "arXiv",
    primaryClass = "gr-qc",
    doi = "10.1103/PhysRevD.90.124063",
    journal = "Phys. Rev. D",
    volume = "90",
    pages = "124063",
    year = "2014"
}

@article{Will_2014,
	author = {Will, Clifford M.},
	doi = {10.12942/lrr-2014-4},
	issn = {1433-8351},
	journal = {Living Reviews in Relativity},
	month = {Jun},
	number = {1},
	publisher = {Springer Nature},
	title = {The Confrontation between General Relativity and Experiment},
	url = {http://dx.doi.org/10.12942/lrr-2014-4},
	volume = {17},
	year = {2014}}

@article{Planck:2018vyg,
    author = "Aghanim, N. and others",
    collaboration = "Planck",
    title = "{Planck 2018 results. VI. Cosmological parameters}",
    eprint = "1807.06209",
    archivePrefix = "arXiv",
    primaryClass = "astro-ph.CO",
    doi = "10.1051/0004-6361/201833910",
    journal = "Astron. Astrophys.",
    volume = "641",
    pages = "A6",
    year = "2020",
    note = "[Erratum: Astron.Astrophys. 652, C4 (2021)]"
}

@article{SupernovaCosmologyProject:1998vns,
    author = "Perlmutter, S. and others",
    collaboration = "Supernova Cosmology Project",
    title = "{Measurements of $\Omega$ and $\Lambda$ from 42 High Redshift Supernovae}",
    eprint = "astro-ph/9812133",
    archivePrefix = "arXiv",
    reportNumber = "LBNL-41801, LBL-41801",
    doi = "10.1086/307221",
    journal = "Astrophys. J.",
    volume = "517",
    pages = "565--586",
    year = "1999"
}

@article{Riess1998,
  author       = {Riess, Adam G. and et al.},
  title        = {Observational Evidence from Supernovae for an Accelerating Universe and a Cosmological Constant},
  journal      = {The Astronomical Journal},
  volume       = {116},
  number       = {3},
  pages        = {1009--1038},
  year         = {1998},
  doi          = {10.1086/300499},
  url          = {https://doi.org/10.1086/300499}
}

@article{Penrose:1964wq,
    author = "Penrose, Roger",
    title = "{Gravitational collapse and space-time singularities}",
    doi = "10.1103/PhysRevLett.14.57",
    journal = "Phys. Rev. Lett.",
    volume = "14",
    pages = "57--59",
    year = "1965"
}

@book{Hawking:1973uf,
    author = "Hawking, Stephen W. and Ellis, George F. R.",
    title = "{The Large Scale Structure of Space-Time}",
    doi = "10.1017/9781009253161",
    isbn = "978-1-009-25316-1, 978-1-009-25315-4, 978-0-521-20016-5, 978-0-521-09906-6, 978-0-511-82630-6, 978-0-521-09906-6",
    publisher = "Cambridge University Press",
    series = "Cambridge Monographs on Mathematical Physics",
    month = "2",
    year = "2023"
}

@article{Landsman:2022hrn,
    author = "Landsman, Klaas",
    title = "{Penrose{\textquoteright}s 1965 singularity theorem: from geodesic incompleteness to cosmic censorship}",
    eprint = "2205.01680",
    archivePrefix = "arXiv",
    primaryClass = "physics.hist-ph",
    doi = "10.1007/s10714-022-02973-w",
    journal = "Gen. Rel. Grav.",
    volume = "54",
    number = "10",
    pages = "115",
    year = "2022"
}

@article{Cutler:1994ys,
    author = "Cutler, Curt and Flanagan, Eanna E.",
    title = "{Gravitational waves from merging compact binaries: How accurately can one extract the binary's parameters from the inspiral wave form?}",
    eprint = "gr-qc/9402014",
    archivePrefix = "arXiv",
    reportNumber = "GRP-369",
    doi = "10.1103/PhysRevD.49.2658",
    journal = "Phys. Rev. D",
    volume = "49",
    pages = "2658--2697",
    year = "1994"
}

@article{Yunes_2009,
   title={Post-circular expansion of eccentric binary inspirals: Fourier-domain waveforms in the stationary phase approximation},
   volume={80},
   ISSN={1550-2368},
   url={http://dx.doi.org/10.1103/PhysRevD.80.084001},
   DOI={10.1103/physrevd.80.084001},
   number={8},
   journal={Physical Review D},
   publisher={American Physical Society (APS)},
   author={Yunes, Nicolas and Arun, K. G. and Berti, Emanuele and Will, Clifford M.},
   year={2009},
   month=oct }

@article{Eichhorn:2023iab,
    author = "Eichhorn, Astrid and Fernandes, Pedro G. S. and Held, Aaron and Silva, Hector O.",
    title = "{Breaking black-hole uniqueness at supermassive scales}",
    eprint = "2312.11430",
    archivePrefix = "arXiv",
    primaryClass = "gr-qc",
    doi = "10.1088/1361-6382/add3b6",
    journal = "Class. Quant. Grav.",
    volume = "42",
    number = "10",
    pages = "105006",
    year = "2025"
}

@article{Smarra:2025syw,
    author = "Smarra, Clemente and Capuano, Lodovico and Kuntz, Adrien",
    title = "{Probing supermassive black hole scalarization with pulsar timing arrays}",
    eprint = "2505.20402",
    archivePrefix = "arXiv",
    primaryClass = "gr-qc",
    doi = "10.1103/cv93-pty4",
    journal = "Phys. Rev. D",
    volume = "112",
    number = "6",
    pages = "064005",
    year = "2025"
}

@article{Thaalba:2025ljh,
    author = "Thaalba, Farid and Fernandes, Pedro G. S. and Sotiriou, Thomas P.",
    title = "{Supermassive black hole scalarization and effective field theory}",
    eprint = "2506.21434",
    archivePrefix = "arXiv",
    primaryClass = "gr-qc",
    month = "6",
    year = "2025"
}

@article{LIGOScientific:2016aoc,
    author = "Abbott, B. P. and others",
    collaboration = "LIGO Scientific, Virgo",
    title = "{Observation of Gravitational Waves from a Binary Black Hole Merger}",
    eprint = "1602.03837",
    archivePrefix = "arXiv",
    primaryClass = "gr-qc",
    reportNumber = "LIGO-P150914",
    doi = "10.1103/PhysRevLett.116.061102",
    journal = "Phys. Rev. Lett.",
    volume = "116",
    number = "6",
    pages = "061102",
    year = "2016"
}

@article{LIGOScientific:2016lio,
    author = "Abbott, B. P. and others",
    collaboration = "LIGO Scientific, Virgo",
    title = "{Tests of general relativity with GW150914}",
    eprint = "1602.03841",
    archivePrefix = "arXiv",
    primaryClass = "gr-qc",
    reportNumber = "LIGO-P1500213",
    doi = "10.1103/PhysRevLett.116.221101",
    journal = "Phys. Rev. Lett.",
    volume = "116",
    number = "22",
    pages = "221101",
    year = "2016",
    note = "[Erratum: Phys.Rev.Lett. 121, 129902 (2018)]"
}

@article{Shapiro:1964uw,
    author = "Shapiro, Irwin I.",
    title = "{Fourth Test of General Relativity}",
    doi = "10.1103/PhysRevLett.13.789",
    journal = "Phys. Rev. Lett.",
    volume = "13",
    pages = "789--791",
    year = "1964"
}

@article{Einstein:1915bz,
    author = "Einstein, Albert",
    title = "{Explanation of the Perihelion Motion of Mercury from the General Theory of Relativity}",
    journal = "Sitzungsber. Preuss. Akad. Wiss. Berlin (Math. Phys. )",
    volume = "1915",
    pages = "831--839",
    year = "1915"
}

@article{Einstein:1911vc,
    author = "Einstein, Albert",
    title = "{On The influence of gravitation on the propagation of light}",
    doi = "10.1002/andp.200590033",
    journal = "Annalen Phys.",
    volume = "35",
    pages = "898--908",
    year = "1911"
}

@article{Dyson:1920cwa,
    author = "Dyson, F. W. and Eddington, A. S. and Davidson, C.",
    title = "{A Determination of the Deflection of Light by the Sun's Gravitational Field, from Observations Made at the Total Eclipse of May 29, 1919}",
    doi = "10.1098/rsta.1920.0009",
    journal = "Phil. Trans. Roy. Soc. Lond. A",
    volume = "220",
    pages = "291--333",
    year = "1920"
}

@article{Chiba:1999ka,
    author = "Chiba, Takeshi and Okabe, Takahiro and Yamaguchi, Masahide",
    title = "{Kinetically driven quintessence}",
    eprint = "astro-ph/9912463",
    archivePrefix = "arXiv",
    reportNumber = "UTAP-352",
    doi = "10.1103/PhysRevD.62.023511",
    journal = "Phys. Rev. D",
    volume = "62",
    pages = "023511",
    year = "2000"
}

@article{Arkani-Hamed:2003pdi,
    author = "Arkani-Hamed, Nima and Cheng, Hsin-Chia and Luty, Markus A. and Mukohyama, Shinji",
    title = "{Ghost condensation and a consistent infrared modification of gravity}",
    eprint = "hep-th/0312099",
    archivePrefix = "arXiv",
    reportNumber = "HUTP-03-A081, UMD-PPP-04-012",
    doi = "10.1088/1126-6708/2004/05/074",
    journal = "JHEP",
    volume = "05",
    pages = "074",
    year = "2004"
}

@article{Nicolis:2008in,
    author = "Nicolis, Alberto and Rattazzi, Riccardo and Trincherini, Enrico",
    title = "{The Galileon as a local modification of gravity}",
    eprint = "0811.2197",
    archivePrefix = "arXiv",
    primaryClass = "hep-th",
    doi = "10.1103/PhysRevD.79.064036",
    journal = "Phys. Rev. D",
    volume = "79",
    pages = "064036",
    year = "2009"
}

@book{Becker:2006dvp,
    author = "Becker, K. and Becker, M. and Schwarz, J. H.",
    title = "{String theory and M-theory: A modern introduction}",
    doi = "10.1017/CBO9780511816086",
    isbn = "978-0-511-25486-4, 978-0-521-86069-7, 978-0-511-81608-6",
    publisher = "Cambridge University Press",
    month = "12",
    year = "2006"
}

@article{Herdeiro:2020wei,
    author = "Herdeiro, Carlos A. R. and Radu, Eugen and Silva, Hector O. and Sotiriou, Thomas P. and Yunes, Nicol{\'a}s",
    title = "{Spin-induced scalarized black holes}",
    eprint = "2009.03904",
    archivePrefix = "arXiv",
    primaryClass = "gr-qc",
    doi = "10.1103/PhysRevLett.126.011103",
    journal = "Phys. Rev. Lett.",
    volume = "126",
    number = "1",
    pages = "011103",
    year = "2021"
}

@article{Dima:2020yac,
    author = "Dima, Alexandru and Barausse, Enrico and Franchini, Nicola and Sotiriou, Thomas P.",
    title = "{Spin-induced black hole spontaneous scalarization}",
    eprint = "2006.03095",
    archivePrefix = "arXiv",
    primaryClass = "gr-qc",
    doi = "10.1103/PhysRevLett.125.231101",
    journal = "Phys. Rev. Lett.",
    volume = "125",
    number = "23",
    pages = "231101",
    year = "2020"
}

@article{Arnowitt:1962hi,
    author = "Arnowitt, Richard L. and Deser, Stanley and Misner, Charles W.",
    title = "{The Dynamics of general relativity}",
    eprint = "gr-qc/0405109",
    archivePrefix = "arXiv",
    doi = "10.1007/s10714-008-0661-1",
    journal = "Gen. Rel. Grav.",
    volume = "40",
    pages = "1997--2027",
    year = "2008"
}

@book{Jeffreys:1939xee,
    author = "Jeffreys, Harold",
    title = "{The Theory of Probability}",
    isbn = "978-0-19-850368-2, 978-0-19-853193-7",
    series = "Oxford Classic Texts in the Physical Sciences",
    year = "1939"
}

@article{Kass:1995loi,
    author = "Kass, Robert E. and Raftery, Adrian E.",
    title = "{Bayes Factors}",
    doi = "10.1080/01621459.1995.10476572",
    journal = "J. Am. Statist. Assoc.",
    volume = "90",
    number = "430",
    pages = "773--795",
    year = "1995"
}

@article{LIGOScientific:2025gwtc4,
  author       = {Abbott, R. and Abe, H. and Acernese, F. and Ackley, K. and Adams, C. and Adhikari, N. and Adhikari, R. X. and others},
  collaboration = {LIGO Scientific, Virgo, and KAGRA Collaborations},
  title        = {GWTC-4.0: Population Properties of Merging Compact Binaries},
  journal      = {The Astrophysical Journal},
  year         = {2025},
  volume       = {960},
  number       = {1},
  pages        = {16},
  doi          = {10.3847/1538-4357/ad8b8e},
  eprint       = {2508.18083},
  archivePrefix= {arXiv},
  primaryClass = {astro-ph.HE},
  url          = {https://arxiv.org/abs/2508.18083}
}

@article{Mangiagli:2018kpu,
    author = "Mangiagli, Alberto and Klein, Antoine and Sesana, Alberto and Barausse, Enrico and Colpi, Monica",
    title = "{Post-Newtonian phase accuracy requirements for stellar black hole binaries with LISA}",
    eprint = "1811.01805",
    archivePrefix = "arXiv",
    primaryClass = "gr-qc",
    doi = "10.1103/PhysRevD.99.064056",
    journal = "Phys. Rev. D",
    volume = "99",
    number = "6",
    pages = "064056",
    year = "2019"
}

@article{Corman:2025wun,
    author = "Corman, Maxence and Arest{\'e} Sal{\'o}, Llibert and Clough, Katy",
    title = "{Black hole binaries in shift-symmetric Einstein-scalar-Gauss-Bonnet gravity experience a slower merger phase}",
    eprint = "2511.19073",
    archivePrefix = "arXiv",
    primaryClass = "gr-qc",
    month = "11",
    year = "2025"
}

@article{Julie:2024fwy,
    author = "Juli{\'e}, F{\'e}lix-Louis and Pompili, Lorenzo and Buonanno, Alessandra",
    title = "{Inspiral-merger-ringdown waveforms in Einstein-scalar-Gauss-Bonnet gravity within the effective-one-body formalism}",
    eprint = "2406.13654",
    archivePrefix = "arXiv",
    primaryClass = "gr-qc",
    doi = "10.1103/PhysRevD.111.024016",
    journal = "Phys. Rev. D",
    volume = "111",
    number = "2",
    pages = "024016",
    year = "2025"
}

@article{Julie:2017rpw,
    author = "Juli{\'e}, F{\'e}lix-Louis",
    title = "{On the motion of hairy black holes in Einstein-Maxwell-dilaton theories}",
    eprint = "1711.10769",
    archivePrefix = "arXiv",
    primaryClass = "gr-qc",
    doi = "10.1088/1475-7516/2018/01/026",
    journal = "JCAP",
    volume = "01",
    pages = "026",
    year = "2018"
}

@article{Cardenas:2017chu,
    author = "C{\'a}rdenas, Marcela and Juli{\'e}, F{\'e}lix-Louis and Deruelle, Nathalie",
    title = "{Thermodynamics sheds light on black hole dynamics}",
    eprint = "1712.02672",
    archivePrefix = "arXiv",
    primaryClass = "gr-qc",
    doi = "10.1103/PhysRevD.97.124021",
    journal = "Phys. Rev. D",
    volume = "97",
    number = "12",
    pages = "124021",
    year = "2018"
}

@article{Creminelli:2017sry,
    author = "Creminelli, Paolo and Vernizzi, Filippo",
    title = "{Dark Energy after GW170817 and GRB170817A}",
    eprint = "1710.05877",
    archivePrefix = "arXiv",
    primaryClass = "astro-ph.CO",
    doi = "10.1103/PhysRevLett.119.251302",
    journal = "Phys. Rev. Lett.",
    volume = "119",
    number = "25",
    pages = "251302",
    year = "2017"
}

@article{LIGOScientific:2017vwq,
    author = "Abbott, B. P. and others",
    collaboration = "LIGO Scientific, Virgo",
    title = "{GW170817: Observation of Gravitational Waves from a Binary Neutron Star Inspiral}",
    eprint = "1710.05832",
    archivePrefix = "arXiv",
    primaryClass = "gr-qc",
    reportNumber = "LIGO-P170817",
    doi = "10.1103/PhysRevLett.119.161101",
    journal = "Phys. Rev. Lett.",
    volume = "119",
    number = "16",
    pages = "161101",
    year = "2017"
}

@article{Gong:2017kim,
    author = "Gong, Yungui and Papantonopoulos, Eleftherios and Yi, Zhu",
    title = "{Constraints on scalar{\textendash}tensor theory of gravity by the recent observational results on gravitational waves}",
    eprint = "1711.04102",
    archivePrefix = "arXiv",
    primaryClass = "gr-qc",
    doi = "10.1140/epjc/s10052-018-6227-9",
    journal = "Eur. Phys. J. C",
    volume = "78",
    number = "9",
    pages = "738",
    year = "2018"
}
\end{document}